\documentclass[12pt]{article}
\usepackage{mathbbol}

\def\beq{\begin{equation}}
\def\eeq{\end{equation}}
\def\bea{\begin{array}}
\def\eea{\end{array}}
\def\beqa{\begin{eqnarray}}
\def\eeqa{\end{eqnarray}}
\def\Pexp{{\rm Pexp}}
\def\cO{{\cal{O}}}

\def\cA{{\cal{A}}}
\def\cN{{\cal{N}}}
\def\cF{{\cal{F}}}
\def\cV{{\cal{V}}}
\def\cR{{\cal{R}}}
\def\cD{{\cal{D}}}
\def\k{{\bf k}}
\def\ff{{\rm{ff}}}
\def\zm{{\rm{zm}}}
\def\ddz{{\frac{d}{dz}}}
\def\Tr{{\rm Tr}}
\def\tr{{\rm tr}}
\def\diag{{\rm diag}}
\def\pl{{{\cal P}_\infty}}

\newcommand{\re}{\relax{\rm I\kern-.18em R}}

\newcommand{\refeq}[1]{\mbox{Eq.~(\ref{eq:#1})}}
\newcommand{\half}{{\scriptstyle{{1\over 2}}}}
\newcommand{\hhalf}{{\scriptstyle{{1/2}}}}
\newcommand{\third}{{\scriptstyle{{1\over 3}}}}
\newcommand{\quart}{{\scriptstyle{{1\over 4}}}}
\newcommand{\veps}{\varepsilon}
\newcommand{\vphi}{\varphi}

\newcommand{\ein}{{\Eins}}

\begin{document}

\hfill\hbox{INLO-PUB-06/03}
\vskip5mm
\begin{center}{\Large\bf Constituent monopoles through the eyes of fermion 
zero-modes}\\[1cm]
{\bf Falk Bruckmann, D\'aniel N\'ogr\'adi} and {\bf Pierre van Baal}\\[3mm]
{\em Instituut-Lorentz for Theoretical Physics, University
of Leiden,\\ P.O.Box 9506, NL-2300 RA Leiden, The Netherlands}
\end{center}

\section*{Abstract}
We use the fermion zero-modes in the background of multi-caloron solutions 
with non-trivial holonomy as a probe for constituent monopoles. We find in 
general indication for an extended structure. However, for well separated
constituents these become point-like. We analyse this in detail for the 
$SU(2)$ charge 2 case, where one is able to solve the relevant Nahm equation 
exactly, beyond the piecewize constant solutions studied previously. 
Remarkably the zero-mode density can be expressed in the high temperature 
limit as a function of the conserved quantities that classify the solutions 
of the Nahm equation. 

\section{Introduction}\label{sec:intro}

To describe regular monopoles in gauge theories a Higgs field is required. 
This de\-fines the abelian subgroup of the gauge field. Yet in the full 
non-abelian theory there is no Dirac string and a regular solution results, 
the well-known 't Hooft-Polyakov monopole~\cite{THPo}. In the strong 
interactions no such Higgs field should be present, but nevertheless it has 
been conjectured that a dual superconductor description, in which monopoles 
form the dual charges that condense, could explain confinement~\cite{DuSC}. 
This scenario receives some support from the studies in supersymmetric 
theories through Seiberg-Witten duality~\cite{SeWi}, although also the old 
center vortex picture is still under active study~\cite{GrRe}. Lattice studies 
based on abelian and center projections, and their respective notions of 
dominance~\cite{Suzu} are the main means through which one tries to address 
these issues. One relies on so-called gauge fixing singularities to identify 
the relevant monopole~\cite{AbPr} or center vortex degrees of freedom.

A more recent alternative to study the monopole content of gauge theories, 
without the need of addressing singular configurations, gauge fixing, nor 
introducing an extra Higgs field, has been through calorons, which are 
instantons at finite temperature. It has been found that calorons are actually 
made up from constituent monopoles~\cite{NCal,Lee,KvB}, which becomes 
most apparent when the background Polyakov loop is non-trivial (as in the 
confined phase), and the size of the caloron is larger than the inverse 
temperature (the extent of the euclidean time direction). The background 
Polyakov loop is defined in the periodic gauge $A_\mu(\vec x,t)=
A_\mu(\vec x,t+\beta)$ by its asymptotic value, or the so-called holonomy, 
\beq
\pl=\lim_{x\to\infty}\Pexp(\int_0^\beta A_0(\vec x,t)dt)=g^\dagger
\exp(2\pi i\diag(\mu_1,\mu_2,\ldots,\mu_n))g,
\eeq
where $g$ is the gauge rotation used to diagonalize $\pl$, whose eigenvalues 
$\exp(2\pi i\mu_j)$ can be ordered on the circle such that $\mu_1\leq\mu_2
\leq\ldots\leq\mu_n\leq\mu_{n+1}$, with $\mu_{n+j}\equiv 1+\mu_j$ and 
$\sum_{i=1}^n\mu_i=0$. The masses of the constituent monopoles are given 
by $8\pi^2\nu_j/\beta$, with $\nu_j\equiv\mu_{j+1}-\mu_j$, which add up 
to $8\pi^2/\beta$, consistent with the instanton action.

Solutions are known explicitly~\cite{Dubna} for $SU(n)$ charge 1 calorons. The 
$n$ constituents can have any spatial location, although all have the same 
location in time (they do, however, become static when well separated), 
and $n-1$ abelian phases complete its $4n$ parameters. Charge $k$ calorons 
can be viewed as composed of $kn$ monopoles, of which a class of axially 
symmetric configurations was constructed explicitly~\cite{BrvB}.

The purpose of this paper is to study in more detail these higher charge 
calorons, where the emphasis is on constructing the chiral fermion 
zero-modes. Charge 1 calorons have exactly one fermion zero-mode, which was 
shown for well separated constituents to be supported on one and only one 
constituent~\cite{MTCP,MTP}. We may change the constituent that supports the
zero-mode, by changing the fermion boundary conditions from (anti)periodic,
to being periodic up to a phase $\exp(2\pi iz)$ (from now on we will use the 
classical scale invariance to set $\beta=1$). For $z\in[\mu_j,\mu_{j+1}]$ the 
zero-mode is localized to what we will call type $j$ constituent monopoles 
(with mass $8\pi^2\nu_j$, and the appropriate $U^{n-1}(1)$ charge associated 
to their embedding in $SU(n)$). 

\begin{figure}[htb]
\vskip2.7cm
\includegraphics{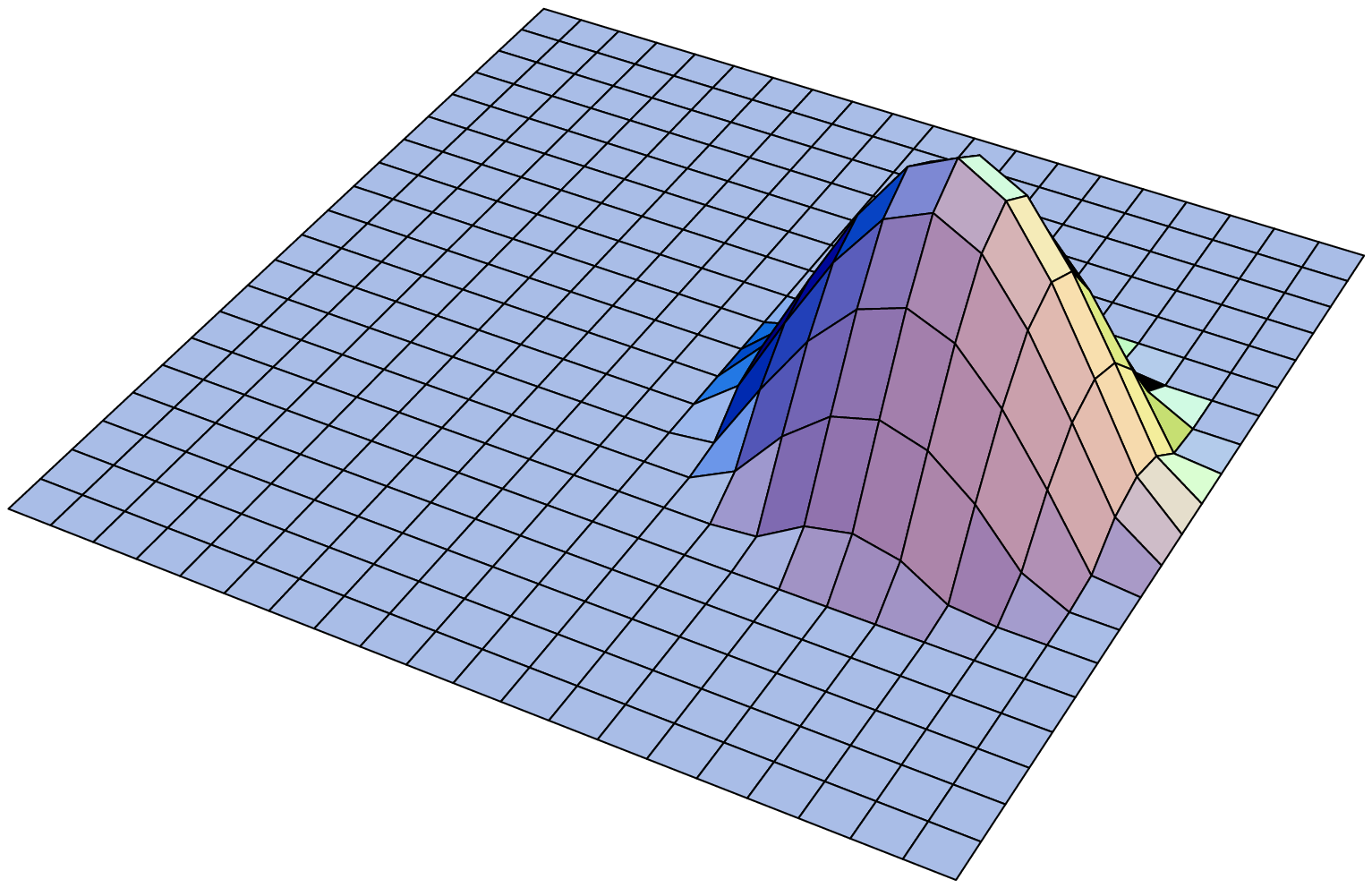}
\includegraphics{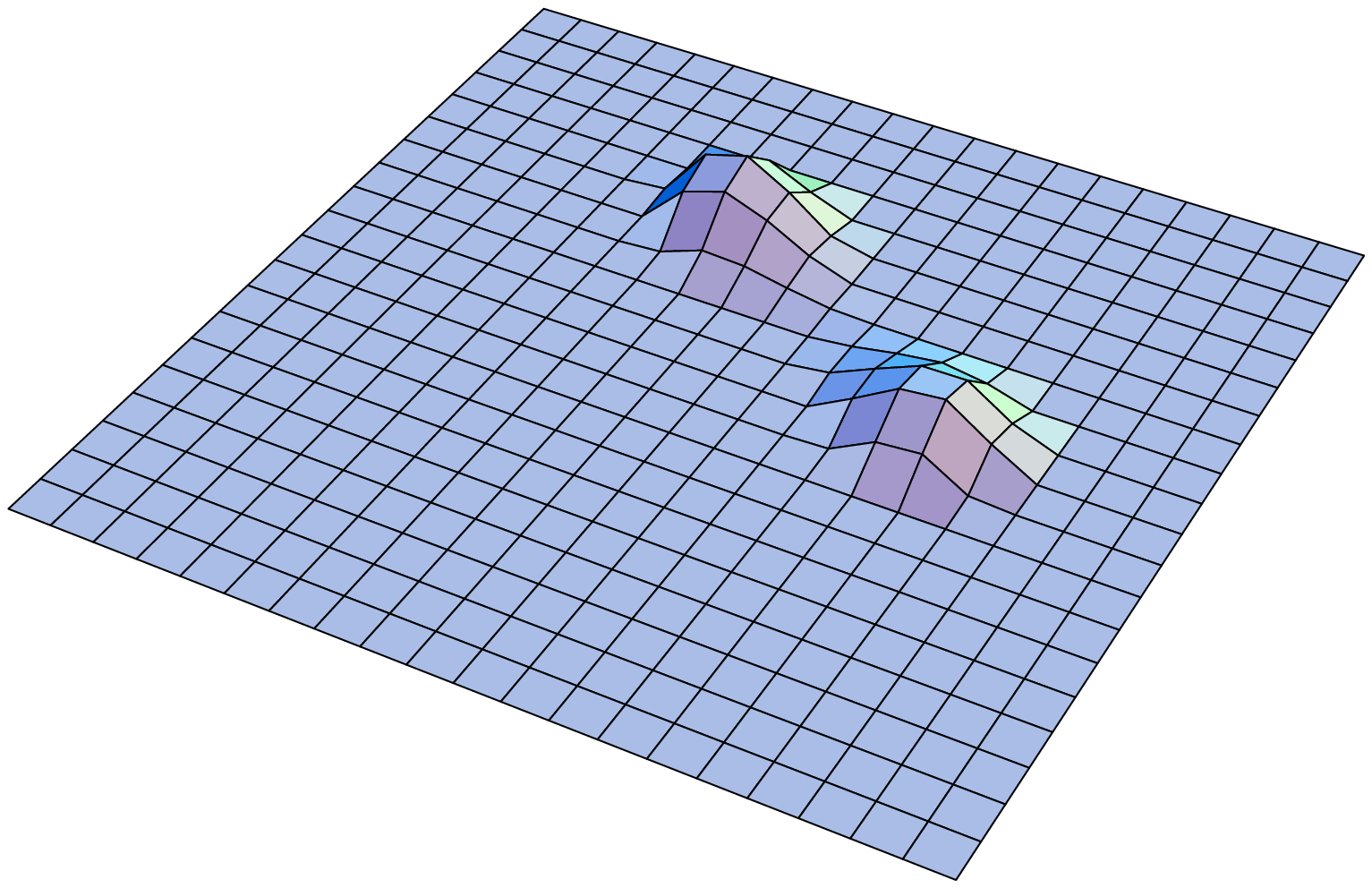}
\includegraphics{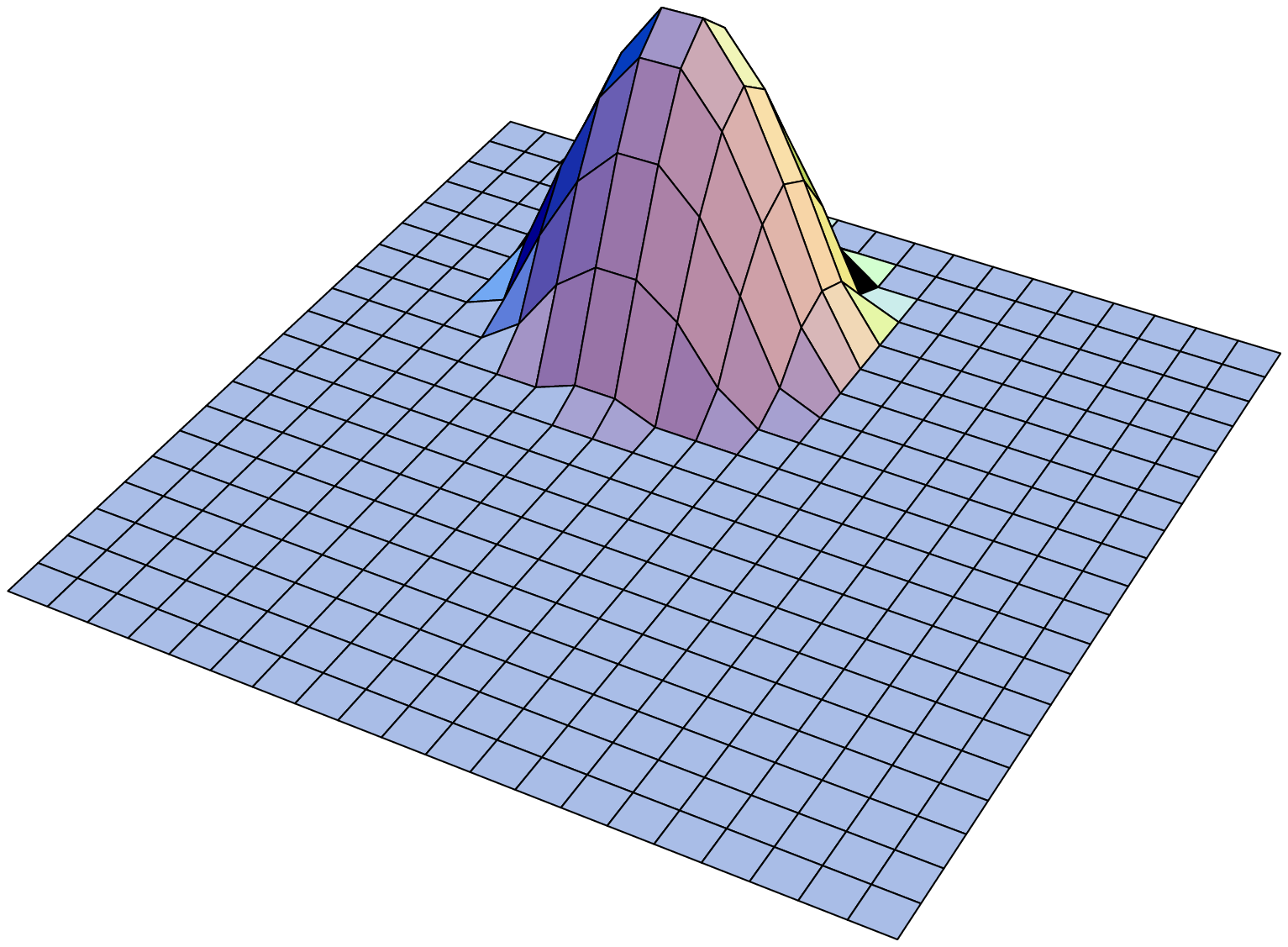}
\hskip0.5cm$z=12/60$\hskip3.1cm$z=19/60$\hskip3.1cm$z=30/60$
\vskip2.7cm
\includegraphics{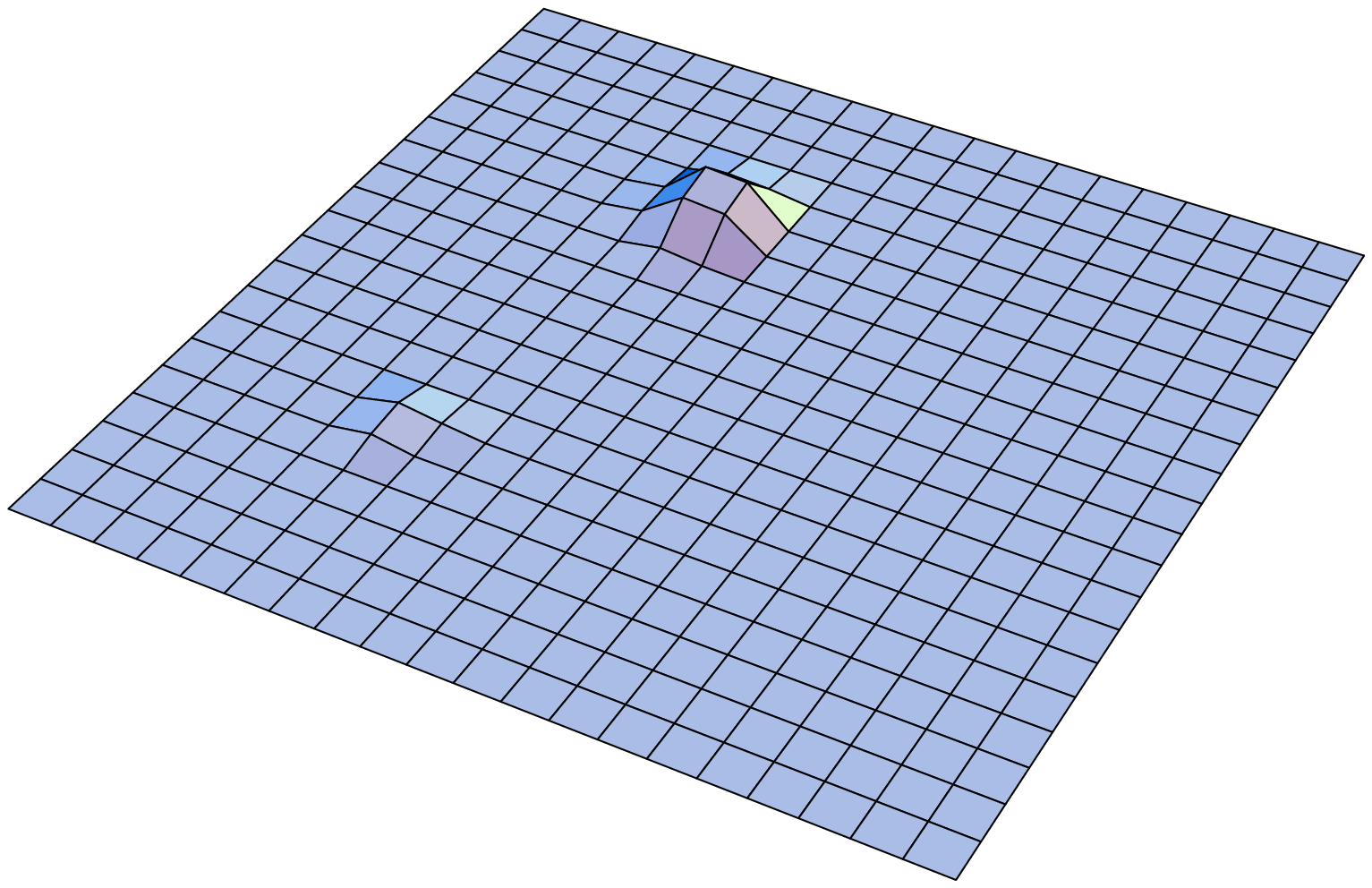}
\includegraphics{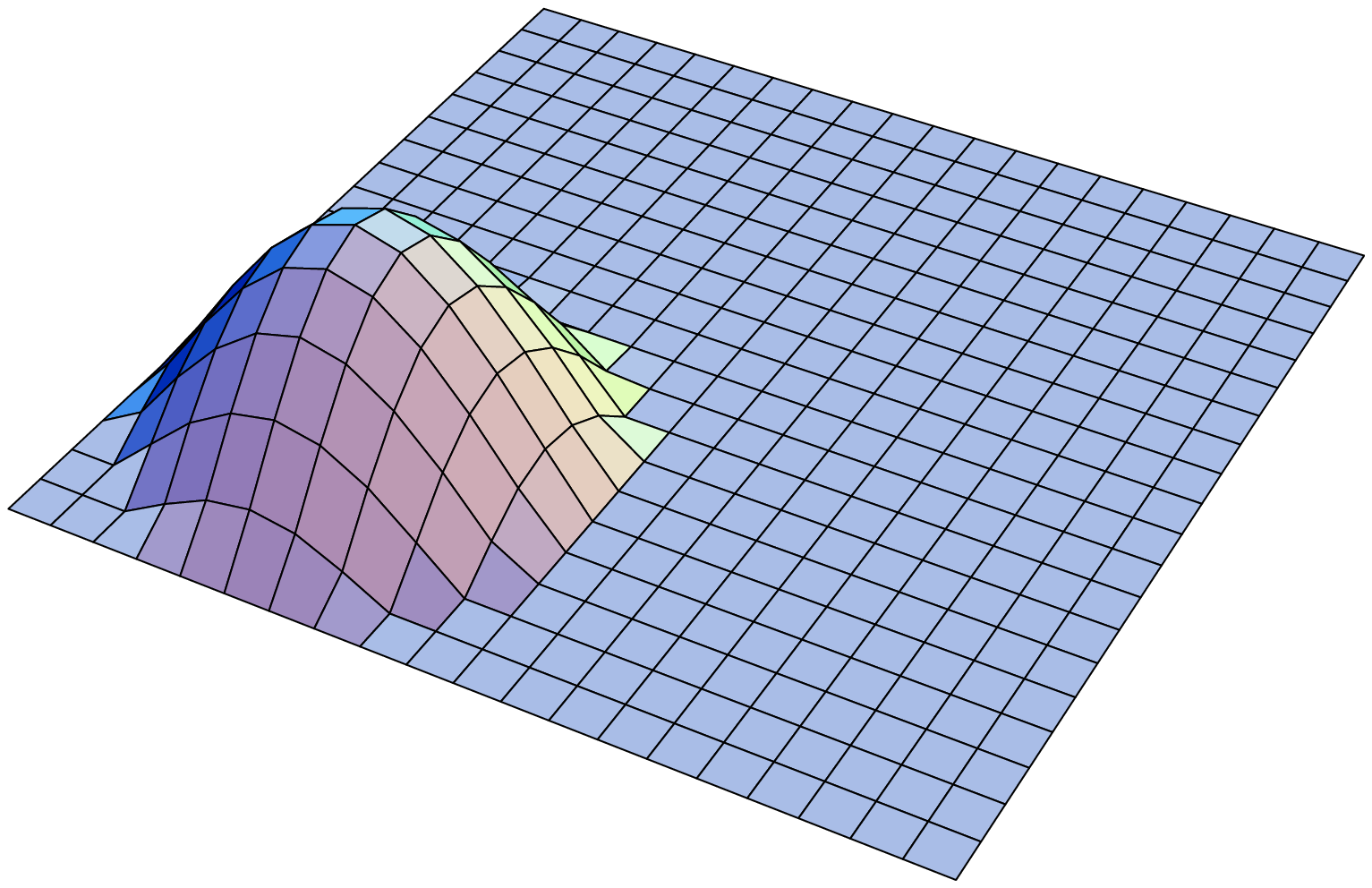}
\includegraphics{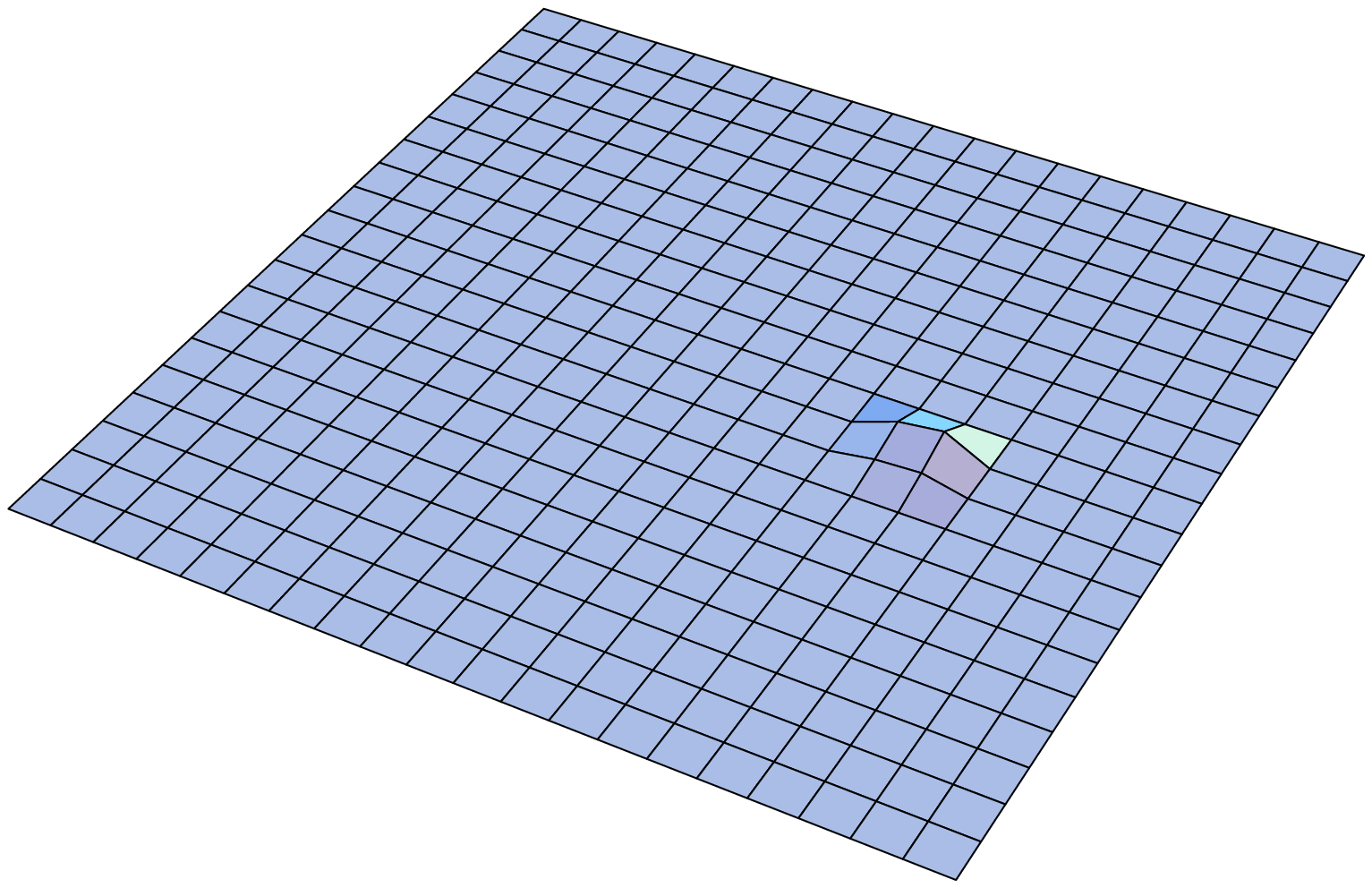}
\hskip0.5cm$z=43/60$\hskip3.1cm$z=48/60$\hskip3.1cm$z=58/60$
\caption{The logarithm of the properly normalized zero-mode density for a 
typical $SU(3)$ caloron of charge 1, cycling through $z$. Shown are $z=\mu_j$ 
(for linear plots see Fig.~\ref{fig:zmrcycle}) and three values of $z$ roughly 
in the middle of each interval $z\in[\mu_j,\mu_{j+1}]$. All plots are on the 
same scale, cutoff for values of the logarithm below -5. The zero-mode with 
anti-periodic boundary conditions is found at $z=30/60$. For the action density
of the associated gauge field, see Ref.~\cite{MTP,WWW}.}\label{fig:zmcycle}
\vskip2.9cm
\includegraphics{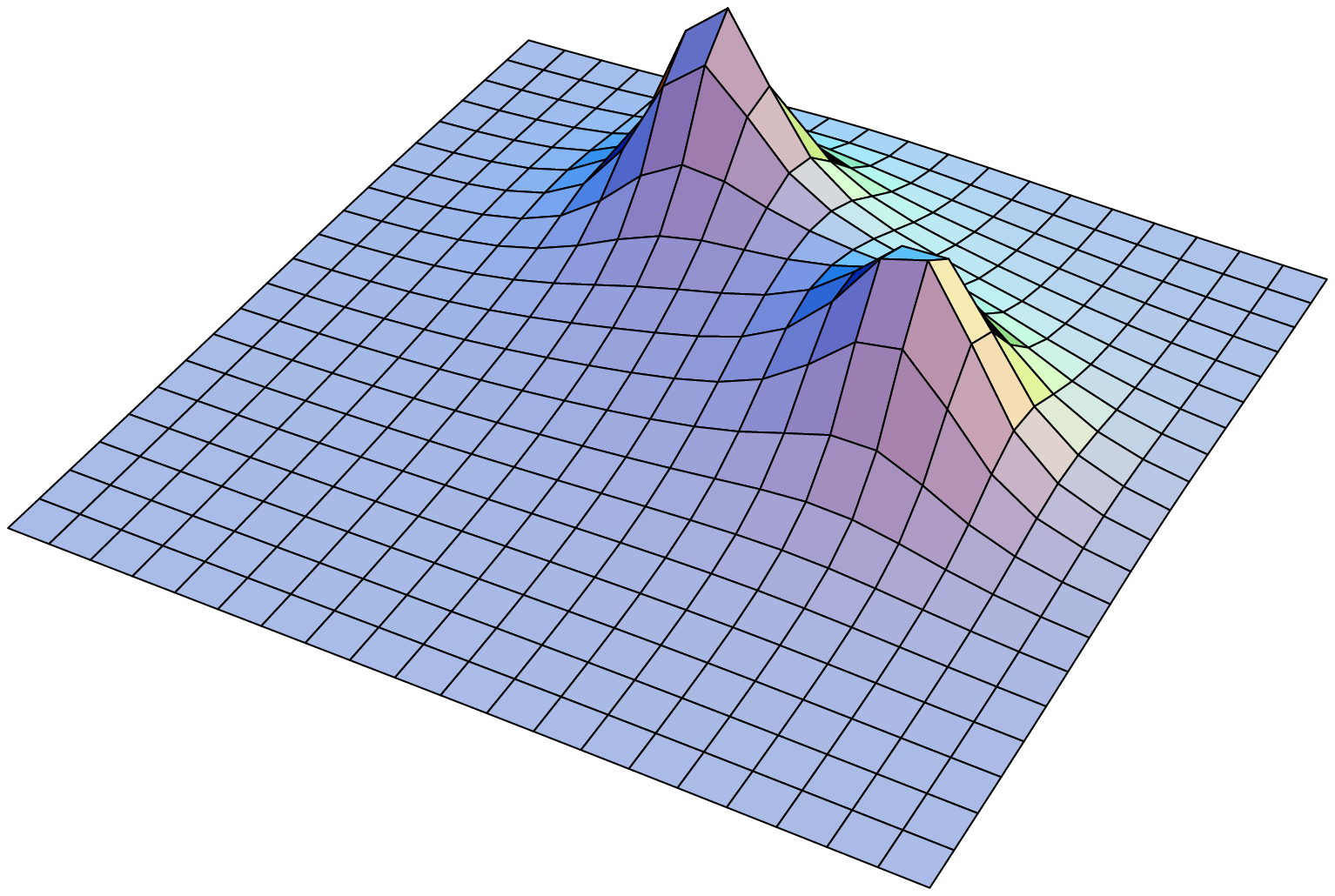}
\includegraphics{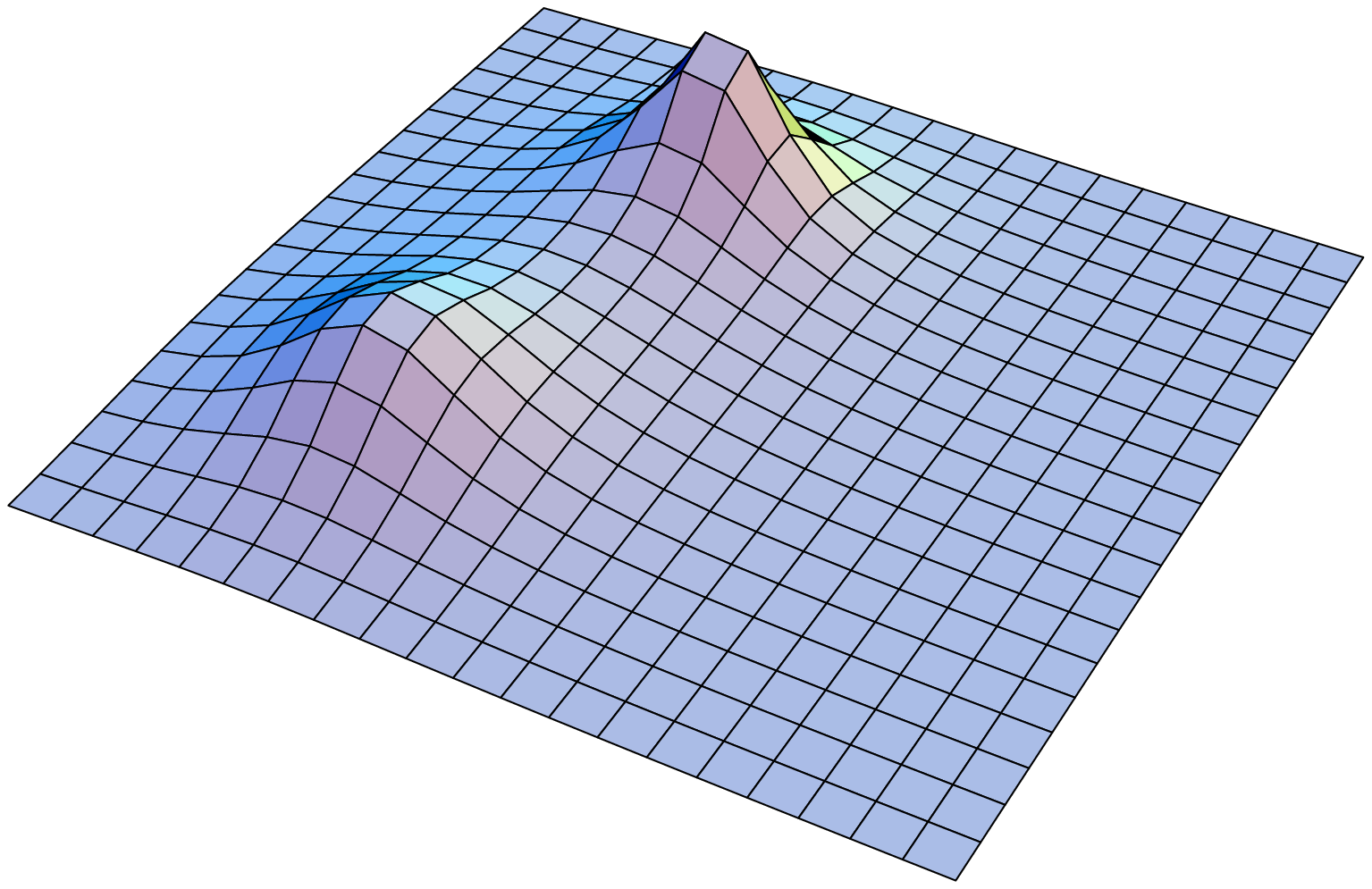}
\includegraphics{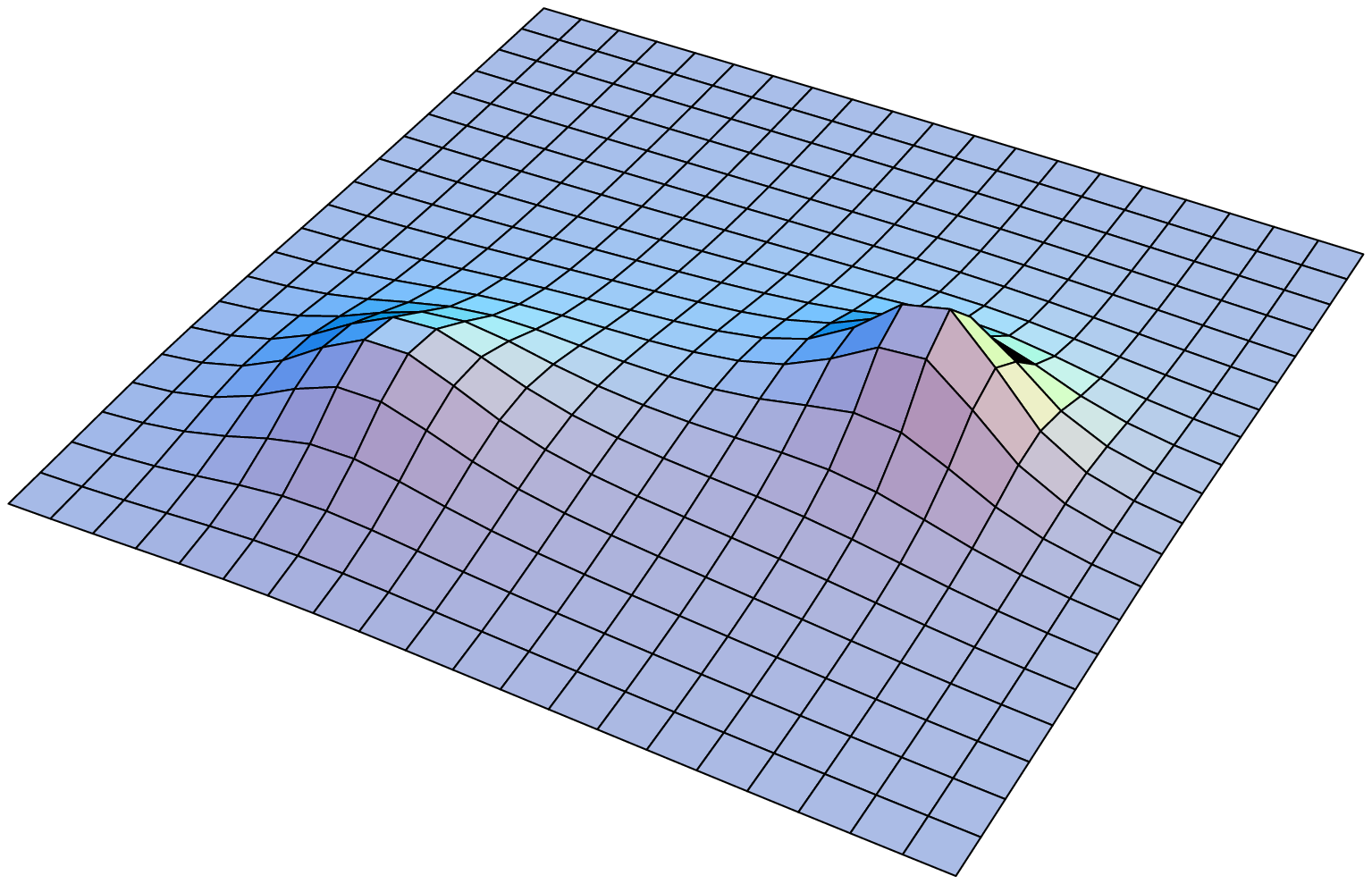}
\hskip0.3cm$z=19/60$\hskip3.1cm$z=43/60$\hskip3.1cm$z=58/60$
\caption{The properly normalized delocalized zero-mode densities for 
$z=\mu_j$, on a {\em linear} scale (cmp. Fig.~\ref{fig:zmcycle}). 
}\label{fig:zmrcycle}
\end{figure}

Lattice evidence has been gathered over recent years that these monopole 
constituents are present in dynamical configurations in the confined phase of 
gauge theories for $SU(2)$ using cooling~\cite{Ilg1,Ilg2}, and for $SU(3)$ 
using fermion zero-modes~\cite{Gattp} as a filter. It is somewhat of a  puzzle 
that these constituent monopoles had not been seen in earlier cooling studies 
(apart from when using twisted boundary conditions~\cite{MTAP}). That they 
remained unnoticed when using fermion zero-modes as a filter is, however, 
simply a consequence of the fact that these studies were restricted to the 
use of fixed fermion boundary conditions. Only when cycling through boundary 
conditions specified by periodicity up to a phase, the $SU(3)$ charge 1 
instanton configurations will show three separate constituent monopoles. 
In Fig.~\ref{fig:zmcycle} we show a typical example based on the exact 
solutions for $SU(3)$, closely following the observed behavior~\cite{Gattp} 
based on actual lattice simulations in the confined phase, which guarantees 
the background Polyakov loop to be non-trivial. In the high temperature 
phase, where the Polyakov loop is trivial, two of the constituents are 
massless and only one peak will be seen. These massless constituents are 
interesting in their own right, giving rise to so-called non-abelian 
clouds~\cite{Wein}, but they will not concern us here.

We do not only study the fermion zero-modes for the higher charge calorons 
to compare with lattice simulations, but also as a tool to understand to 
which extend the constituent monopoles can be unambiguously identified in 
the caloron solutions. In the high temperature limit the non-abelian cores 
of the monopoles shrink to zero-sizes, and one is left with abelian gauge 
fields. Without taking the high temperature limit, but excluding the 
non-abelian cores, the same physics describes what we called the far field 
region~\cite{BrvB}. It would be desirable if the abelian field in this region
is described by point-like Dirac monopoles (actually dyons because of 
self-duality), when the constituents are well separated. For charge 1 calorons 
and the class of axially symmetric solutions studied before~\cite{BrvB} the 
density of the fermion zero-modes become Dirac delta functions at the 
locations of the constituent monopoles in the high temperature limit, 
for any constituent separation. This infinite localization in the high 
temperature limit can be understood from the fact that for most $z$ values 
there is an effective mass for the fermions. Therefore, by studying these 
zero-modes in the general case, we hope to learn more about the localization 
of the monopoles. 

It is not directly obvious that any higher charge caloron can be described 
by point-like constituents in the far field limit. The tool to construct 
these solutions is the Nahm equation~\cite{Nahm,NCal}, which is a duality 
transformation that maps the problem of finding charge $k$ calorons to that 
of $U(k)$ gauge fields on a circle with specific singularities at $z=\mu_j$. 
In the cases studied so far, the dual (Nahm) gauge field can be made piecewize 
constant, from which one easily reads off the constituent monopole locations. 
In general, however, the Nahm gauge fields depend non-trivially on $z$, and it 
is important to understand what this implies for the localization of the 
constituents. Quite remarkably, we will nevertheless find that in the high 
temperature limit the fermion zero-mode density does not depend on $z$ and can 
be expressed in terms of the conserved quantities of the Nahm equation. We 
compute it explicitly for charge 2, revealing in general an extended
structure. However, the extended structure collapses to isolated points for 
well separated constituents. This is so as long as $z\neq\mu_j$, which is the 
value where the localization of the zero-modes jumps. We separately study 
in the high temperature limit the case where $z=\mu_j$, for which the fermion 
zero-modes delocalize (decaying algebraically). Support of the zero-modes is 
now on constituent monopoles of types $j-1$ and $j$. We analyse this in 
detail for the class of axially symmetric solutions of Ref.~\cite{BrvB}, in 
the light of some puzzles concerning so-called bipole zero-modes~\cite{Bip}.

This paper is organized as follows: first we describe the caloron zero-modes 
in Sect.~\ref{sec:zm} and set up the Green's function calculation in 
Sect.~\ref{sec:zmlim}, to be able to discuss the zero-mode and far field 
limits. Sect.~\ref{sec:bipole} deals with the case where $z=\mu_j$, for which 
zero-modes are delocalized. In Sect.~\ref{sec:cons} we relate the zero-mode 
density for $z\neq\mu_j$ to conserved quantities of the Nahm equation and 
study in Sect.~\ref{sec:chtwo} the general solution for $SU(2)$ charge 2 
caloron. We conclude with a discussion of the implications of our results 
and the problems that still need to be addressed. Two appendices provide the 
details for the zero-mode and far field limit calculations. 

\section{Fermion zero-modes}\label{sec:zm}

We wish to construct the zero-modes of the chiral Dirac equation in 
the background of a self-dual gauge field at finite temperature. 
The Dirac equation in its two-component Weyl form, with $\bar
\sigma_\mu=\sigma_\mu^\dagger=(\ein_2,-i\vec \tau)$ ($\tau_i$ are the
usual Pauli matrices), reads
\beq
\bar D\hat\Psi_z(x)\equiv\bar\sigma_\mu D_\mu\hat\Psi_z(x)\equiv
\bar\sigma_\mu(\partial_\mu+A_\mu(x))\hat\Psi_z(x).\label{eq:Dirac}
\eeq
Assuming the gauge field is periodic in the imaginary time direction, with 
period $\beta=1$, we seek the zero-modes that satisfy the boundary condition 
\beq
\hat\Psi_z(t+1,\vec x)=\exp(-2\pi iz)\hat\Psi_z(t,\vec x),
\eeq
A simple abelian gauge transformation, $\Psi_z(x)=\exp(2\pi i zt)
\hat\Psi_z(x)$, makes the zero-mode periodic. This gauge transformation
replaces the gauge field by $A_\mu(x)-2\pi i z_\mu\ein_n$, but it does 
not change the field strength, such that existence of the appropriate 
number of zero-modes is guaranteed by the index theorem. 

The caloron solutions are obtained by Fourier transformation, reformulating 
the algebraic ADHM (Atiyah-Drinfeld-Hitchin-Manin) construction~\cite{ADHM} 
of multi-instanton solutions in $R^4$, as the Nahm transformation~\cite{NCal}. 
For this the instantons in $R^4$ periodically repeat themselves in the 
imaginary time direction up to a gauge rotation with $\pl$. This Fourier 
transformation also selects out of the infinite number of fermion zero-modes 
in $R^4$, those that satisfy the correct periodicity.

In the ADHM formulation the $k$ normalized fermion zero-modes for a charge $k$ 
$SU(N)$ instanton are given by ($\veps\equiv i\tau_2=\sigma_2$, $i=1,\ldots,n$ 
is the color index, $l=1,\ldots,k$ the charge index and $I=1,2$ the spinor 
index)
\beq
\Psi_{iI}^l(x)=\pi^{-1}\left(\phi^{-\hhalf}(x) u^\dagger(x)f_x\veps
\right)_{iI}^l,\quad\phi(x)=\ein_n+u^\dagger(x)u(x),\label{eq:Psi}
\eeq
with $u^\dagger(x)$ given explicitly in terms of the ADHM parameters by 
\beq
u^\dagger(x)=\lambda(B-\ein_k x)^{-1},\quad B=\sigma_\mu B_\mu,
\quad x=x_\mu\sigma_\mu,
\eeq
where $\lambda=(\lambda_1,\ldots,\lambda_k)$, with $\lambda_i^\dagger$ a 
two-component spinor in the $\bar n$ representation of $SU(n)$ ($\lambda$ 
can be seen as a $n\times 2k$ complex matrix), and $B_\mu$ hermitian 
$k\times k$ matrices. As $\phi(x)$ is an $n\times n$ positive hermitian 
matrix (for $n=2$ proportional to $\ein_2$), its square root is well-defined. 
We also recall the gauge field is given by
\beq
A_\mu(x)=\phi^{-\hhalf}(x)(u^\dagger(x)\partial_\mu u(x))\phi^{-\hhalf}(x)
+\phi^{\hhalf}(x)\partial_\mu\phi^{-\hhalf}(x).
\label{eq:aadhm}
\eeq
which can be shown to be self-dual provided the quadratic ADHM constraint is 
satisfied,
\beq
\lambda^\dagger\lambda+(B-\ein_k x)^\dagger(B-\ein_k x)=\sigma_0 f^{-1}_x,
\label{eq:adhmconstr}
\eeq
which implicitly defines $f_x$ as a hermitian $k\times k$ Green's function,
thereby completing the description of \refeq{Psi}. A further 
simplication~\cite{KvB,Temp} will be helpful, which uses the fact 
that $2(B_\mu-x_\mu)=\partial_\mu f_x^{-1}$ and $u^\dagger(x)=\phi(x)
\lambda f_x(B-\ein_k x)^\dagger$, implying
\beqa
A_\mu(x)&=&\half\phi^\hhalf(x)\lambda\bar\eta_{\mu\nu}\partial_\nu f_x
\lambda^\dagger\phi^\hhalf(x)+\half[\phi^{-\hhalf}(x),\partial_\mu
\phi^\hhalf(x)],\nonumber\\ \phi^{-1}(x)&=&1-\lambda f_x\lambda^\dagger,
\quad\Psi_{iI}^l(x)=(2\pi)^{-1}(\phi^\hhalf(x)\lambda\partial_\mu 
f_x\bar\sigma_\mu\veps)_{iI}^l,\label{eq:defA}
\eeqa
with $\bar\eta_{\mu\nu}=\bar\eta^j_{\mu\nu}\sigma_j=\bar\sigma_{[\mu}
\sigma_{\nu]}$ the anti-selfdual 't Hooft tensor.

As mentioned above, calorons are obtained by arranging the instantons in 
$R^4$ to be periodic (up to a gauge rotation). The time interval $[0,1]$ 
will contain as many instantons as the topological charge of the caloron. 
One splits the charge index $l$ as $l=pk+a$, where $a$ labels the $k$ 
instantons in the interval $[0,1]$ and $p$ labels the infinite number of 
repeated time-intervals, playing the role of the Fourier index. We find, 
suppressing the gauge and spinor index (cmp. Ref.~\cite{MTCP,MTP}), 
\beq
\hat\Psi_z^a(x)=(2\pi)^{-1}\phi^\hhalf(x)\partial_\mu\int_0^1 dz'
\hat\lambda_b(z')\hat f^{ba}_x(z',z)\bar\sigma_\mu\veps.\label{eq:Psihat}
\eeq
where the Fourier transforms of $\lambda$ and $f_x$ are denoted by 
$\hat\lambda_a(z)$ and $\hat f_x^{ba}(z',z)$.
The fermion zero-modes thus constructed are in the so-called algebraic gauge, 
for which $\hat\Psi_z(t+1,\vec x)=\exp(-2\pi iz)\pl\hat\Psi_z(t,\vec x)$. In 
this gauge all components of $A_\mu$ vanish at spatial infinity and the 
non-trivial holonomy is encoded in the boundary condition, which for the 
gauge field reads $A_\mu(t+1,\vec x)=\pl A_\mu(t,\vec x){\cal P}_\infty^{-1}$. 
A simple time-dependent gauge transformation allows one to transform to the 
periodic gauge, after which $A_0$ goes to a constant at spatial infinity.

To encode the appropriate periodicity in the ADHM parameters we need to take 
$\lambda_{pk+a}={\cal P}_\infty^p\zeta_a$~\cite{BrvB}. Introducing the 
projections $P_m$ on the eigenvalues of $\pl$, i.e. $\pl\equiv\sum_{m=1}^n
\exp(2\pi i\mu_m)P_m$, we find $\hat\lambda_a(z)=\sum_{m=1}^n\delta(z-\mu_m)
P_m\zeta_a$, which makes the expression for the zero-modes particularly simple,
\beq
\hat\Psi_z^a(x)=(2\pi)^{-1}\phi^\hhalf(x)\sum_{m=1}^n P_m\zeta_b
\bar\sigma_\mu\veps\partial_\mu\hat f^{ba}_x(\mu_m,z).\label{eq:zm}
\eeq
The fact that we are dealing with higher charge calorons, is reflected in 
the presence of the indices $a,b=1,\ldots,k$. Making use of the well-known 
identity~\cite{Temp,Osb,CoG} in $R^4$, $\Psi_{iI}^l(x)^*\Psi_{iI}^m(x)=
-(2\pi)^{-2}\partial_\mu^2 f^{lm}_x$, Fourier transformation gives the
appropriate expression for the caloron zero-mode density
\beq
\hat\Psi_z^a(x)^\dagger\hat\Psi_z^b(x)=-(2\pi)^{-2}
\partial_\mu^2\hat f^{ab}_x(z,z).\label{eq:zdens}
\eeq
Using the fact that $\lim_{|\vec x|\rightarrow\infty}|\vec x|\hat 
f^{ab}_x(z,z)=\pi\delta^{ab}$, the zero-modes $\hat\Psi_z^a$ are seen 
to be orthonormal.

We close this section by remarking that for $SU(2)$ an alternative 
construction is possible, as part of the $Sp(n)$ series (since $Sp(1)=
SU(2)$). The ADHM construction for $Sp(n)$ is based on quaternions. 
In particular $\lambda_l$ is assumed to be a quaternion, whereas $B_\mu$ 
is now real-symmetric. All formulae presented above remain valid, but it 
should be noted that the transformation $\lambda\rightarrow\lambda 
T^\dagger$, $B_\mu\rightarrow T B_\mu T^\dagger$, with $T\in U(k)$, 
leaving the gauge field and the ADHM constraint untouched, has to 
be replaced by $T\in O(k)$.

\section{Zero-mode and far field limit}\label{sec:zmlim}

As we have seen above, all physical quantities can be reconstructed, once 
we have found the Green's function $\hat f^{ab}_x(z,z')$. Here we review the
necessary ingredients. We start with the fact that the Green's function is 
defined through an impurity scattering problem~\cite{BrvB}
\beq
\left\{-\frac{d^2}{dz^2}+V(z;\vec x)\right\}
\!f_x(z,z')=4\pi^2 \ein_k\delta(z\!-\!z'),\label{eq:Green}
\eeq
where $f_x(z,z')$ is related to $\hat f_x(z,z')$ by a $U(k)$ gauge 
transformation 
\beq
f_x(z,z')=\hat g(z)\hat f_x(z,z')\hat g^\dagger(z'),\quad
\hat g(z)=\exp\left(2\pi i(\xi_0-x_0\ein_k)z\right).\label{eq:dftrans}
\eeq
The ``potential" $V$, which includes ``impurity" contributions,
is determined by the (dual) $U(k)$ Nahm gauge field $\hat A_\mu(z)$ 
\beqa
&&V(z;\vec x)=4\pi^2\vec R^2(z;\vec x)+2\pi\sum_m\delta(z-\mu_m)S_m,\quad
S_m=\hat g(\mu_m)\hat S_m\hat g^\dagger(\mu_m),\nonumber\\&&R_j(z;\vec x)
=x_j\ein_k-(2\pi i)^{-1}\hat g(z)\hat A_j(z)\hat g^\dagger(z),\qquad\,
\hat S_m^{ab}=\pi\tr_2(\zeta_a^\dagger P_m\zeta_b).\label{eq:Rdef}
\eeqa
Fourier transformation of $B_\mu$ gives $(2\pi i)^{-1}\delta(z-z')(
\delta_{\mu0}\ein_k\ddz+\hat A_\mu(z))$ and defines the Nahm gauge field. 
We have further used the fact that one can choose a $U(k)$ gauge in which 
$\hat A_0(z)\equiv 2\pi i\xi_0$ is constant, which itself can be transformed 
to zero by $\hat g(z)$. Note that $\hat g(1)$ plays the role of the holonomy 
associated to the dual Nahm gauge field $\hat A_\mu(z)$. Crucial is that we 
can formulate the zero-mode and far field limits without specifying the 
solutions of the Nahm equations. These Nahm equations, which are equivalent 
to the ADHM constraint, are given by 
\beqa
&&\ddz\hat A_j(z)+[\hat A_0(z),\hat A_j(z)]+\half\veps_{jk\ell}[\hat A_k(z),
\hat A_\ell(z)]=2\pi i\sum_m\delta(z-\mu_m)\rho_m^{\,j},\nonumber\\
&&\hskip4cm\vec\rho_m^{\,ab}\equiv-\pi\tr_2\left(\zeta_a^\dagger P_m\zeta_b
\vec\tau\right).\label{eq:nahm}
\eeqa
This will be discussed in more detail in Sect.~\ref{sec:chtwo}.

To solve the second order equation for $f_x(z,z')$ it is convenient to convert 
it to a first order equation, involving $2k\times 2k$ matrices, 
\beq
\left(\ddz-\pmatrix{0&\ein_k\cr V(z;\vec x)&0\cr}\right)\pmatrix{\quad 
f_x(z,z')\cr\ddz f_x(z,z')\cr}=-4\pi^2\delta(z-z')\pmatrix{0\cr\ein_k\cr},
\eeq
which can be solved as
\beq
\pmatrix{\quad f_x(z,z')\cr\ddz f_x(z,z')\cr}=-4\pi^2 W(z,z_0)\left\{
(\ein_{2k}-\cF_{z_0})^{-1}-\theta(z'-z)\ein_{2k}\right\}W^{-1}(z',z_0)
\pmatrix{0\cr\ein_k\cr},\label{eq:fdf}
\eeq
where $z_0$ can be arbitrary and 
\beq
W(z_2,z_1)=\Pexp\left[\int_{z_1}^{z_2}\pmatrix{0&\ein_k\cr V(z;\vec x)&0\cr}
dz\right],\quad\cF_{z_0}=\hat g^\dagger(1)W(z_0+1,z_0).\label{eq:Wdef}
\eeq
In particular, one can show that~\cite{BrvB} 
\beq
-\half\Tr_n F_{\mu\nu}^2(x)=-\half\partial_\mu^2\partial_\nu^2\log 
\psi(x),\quad\psi=\det\left(ie^{-\pi ix_0}(\ein_{2k}-\cF_{z_0})
/\sqrt{2}\right).\label{eq:defpsi}
\eeq
To isolate the exponential contributions in \refeq{Wdef}, one introduces two 
solutions of the Riccati equation~\cite{BrvB},
\beq
R_m^\pm(z)^2\pm\frac{1}{2\pi}\frac{d}{dz}\,R_m^\pm(z)=\vec
R^2(z;\vec x).\label{eq:Riccati} 
\eeq
Since
$\vec R(z;\vec x)\to\vec x\ein_k$ for $|\vec x|\to\infty$, we find in this
limit that $R_m^\pm(z)\to|\vec x|\ein_k$. Defining
\beq 
f_m^\pm(z)=P\exp\left[\pm 2\pi\int_{\mu_m}^z \!\!\!R_m^\pm(z)dz\right],
\quad z\in[\mu_m,\mu_{m+1}],\label{eq:fdef}
\eeq
we see that $f_m^\pm(z)\to\exp\left(\pm 2\pi|\vec x|(z-\mu_m)\ein_k\right)$.
These are the exponentially rising and decaying solutions of \refeq{Green},
in terms of which we can rewrite for $z,z'\in(\mu_m,\mu_{m+1})$
\beq
W(z,z')=W_m(z)W_m^{-1}(z'),\quad W_m(z)\equiv\hat W_m(z)F_m(z),
\eeq
where $F_m(z)$ contains all exponential factors, 
\beq
\hat W_m(z)=\pmatrix{\ein_k&\ein_k\cr 2\pi R^+_m(z)&-2\pi R^-_m(z)\cr},
\quad F_m(z)=\pmatrix{f_m^+(z)&0\cr0&f_m^-(z)\cr}.\label{eq:Wmdef}
\eeq

As an illustration let us assume $\vec R(z;\vec x)^2=(\vec x\ein_k-\vec e\,
Y_m)^2\equiv R^2_m$, independent of $z$, for $z\in[\mu_m,\mu_{m+1}]$, as is 
the case for charge one~\cite{KvB} and a class of axially symmetric 
solutions~\cite{BrvB} (see Sect.~\ref{sec:bipole}). We then find $R^+_m(z)=
R^-_m(z)=R_m$ and $f_m^\pm(z)=\exp(2\pi(z-\mu)R_m)$. Diagonalizing $R_m$ 
defines $k$ locations (the eigenvalues of $Y_m$), which are sharply defined 
and give rise to point-like constituents in the high temperature limit. When 
$\vec R(z;\vec x)$ is not piecewize constant, these locations are a priori not 
sharply defined. We still expect the cores to be the regions in $\vec x$ where 
$\vec R(z;\vec x)$ remains small (cmp. \refeq{Rdef}). The separations between 
the cores of monopoles of different type is controlled by the discontinuity in 
$\hat A_j(z)$, \refeq{nahm}, and can in general be chosen large. This allows 
us to define the zero-mode limit, where $\vec x$ is assumed to be far removed 
from any constituent core {\em not} of type $m$. In technical terms that means 
that $R_{m'}^\pm(z)$ is large for all $m'\neq m$. Accordingly, $f_{m'}^-(z)$ 
and $f_{m'}^+(z)^{-1}$ are exponentially small for these values of $\vec x$ 
(cmp. \refeq{fdef}). Results that are valid up to these exponential correction 
are denoted by a subscript ``$\zm$" for the zero-mode limit and ``$\ff$" for 
the far field limit. In the latter case, $\vec x$ is assumed to be far removed 
from {\em all} constituents.

\begin{figure}[htb]
\vspace{4.4cm}
\includegraphics{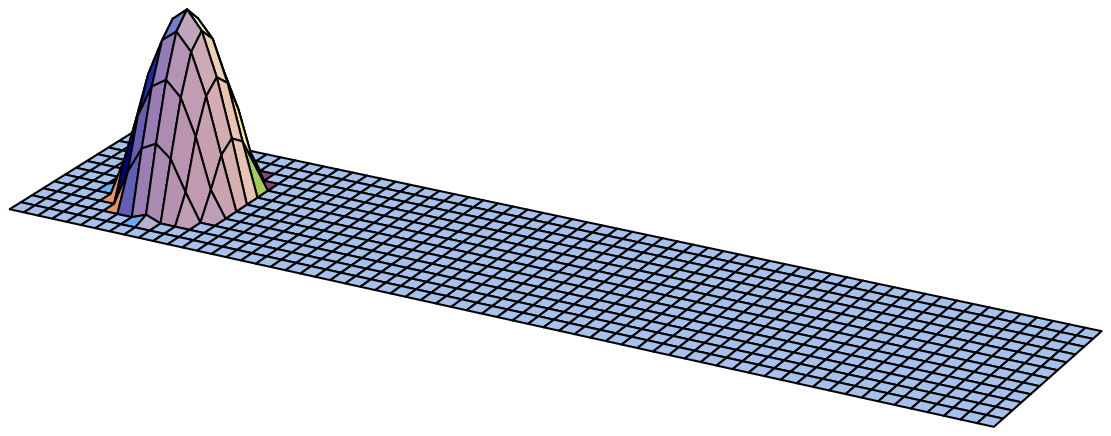}
\includegraphics{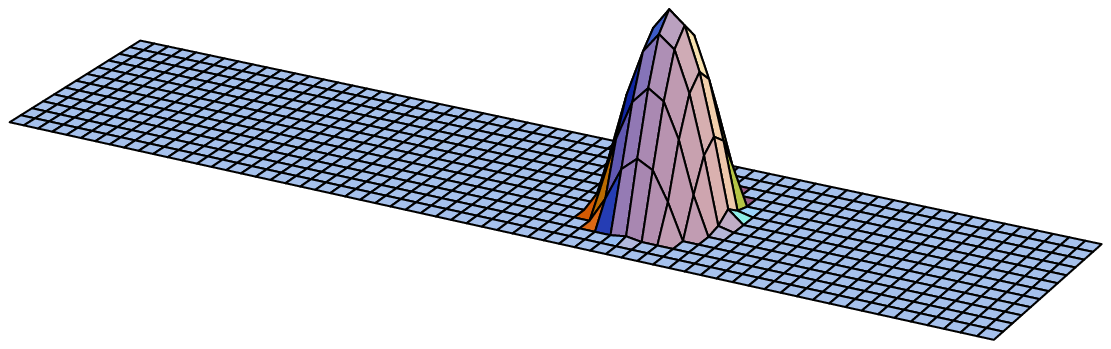}
\includegraphics{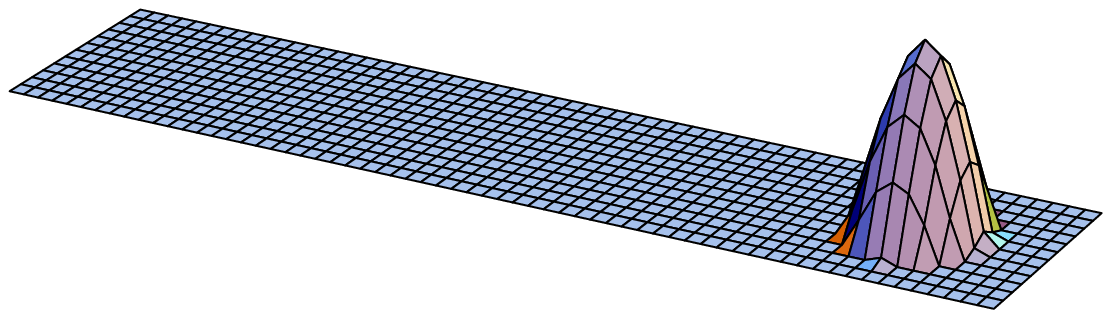}
\includegraphics{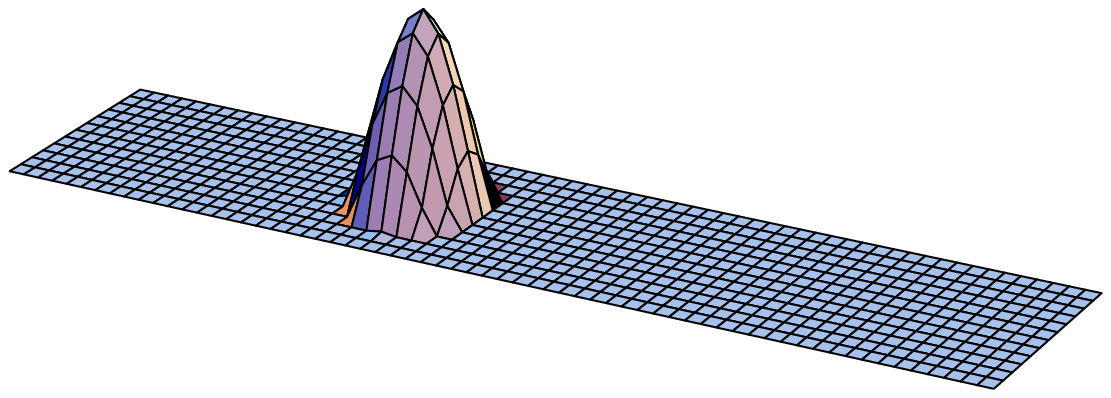}
\includegraphics{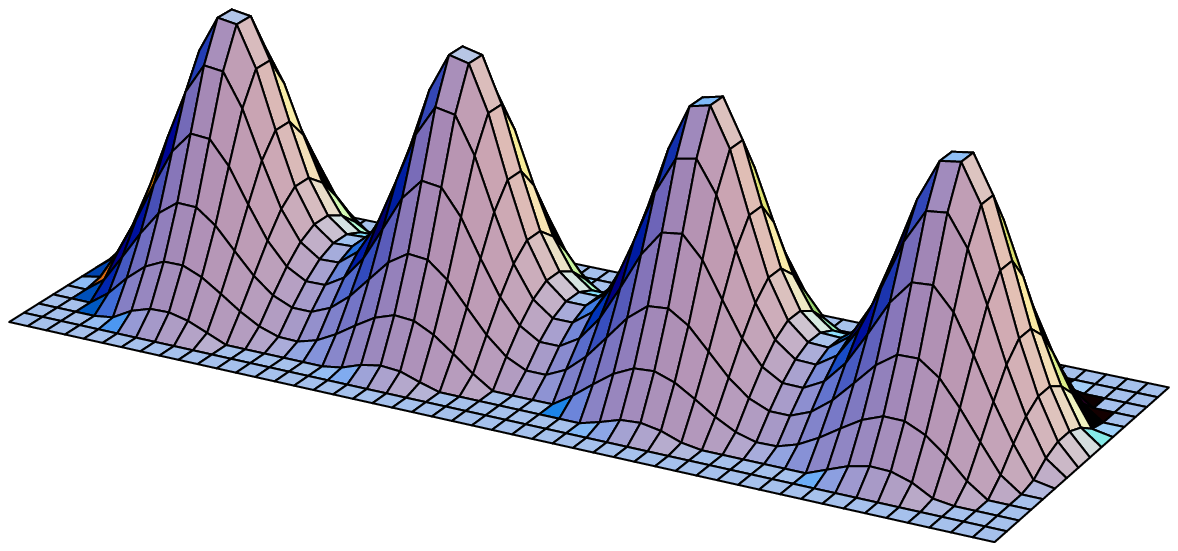}
\caption{Zero-mode densities for a typical charge 2, $SU(2)$ axially symmetric 
solution. For comparison the action density (cmp. Fig.~2 of Ref.~\cite{BrvB})
is shown in the middle. All are on a logarithmic scale, cutoff below $e^{-3}$. 
On the left is shown the two periodic zero-modes ($z=0$) and on the right 
the two anti-periodic zero-modes ($z=1/2$). 
 }\label{fig:zmlocal}
\end{figure}

In appendix A we show that for $\mu_m\leq z'\leq z\leq\mu_{m+1}$ 
\beq
f^{\zm}_x(z,z')=\pi(e_m^+(z)-e_m^-(z)Z_{m+1}^-)\left(e_m^+-Z_m^+e_m^-
Z_{m+1}^-\right)^{-1}(\tilde e_m^+(z')-Z_m^+\tilde e_m^-(z'))R_m^{-1}(z').
\label{eq:zmlim}
\eeq
(for $z'>z$ one uses $f_x(z',z)=f_x^\dagger(z,z')$) with
\beqa
e_m^\pm(z)\equiv f_m^\mp(z)f_m^\mp(\mu_{m+1})^{-1},&&e_m^\pm\equiv 
e_m^\pm(\mu_m),\quad\tilde e_m^\pm(z)\equiv f_m^\mp(z)^{-1},\nonumber\\
Z_m^-\equiv\ein_k-2\Sigma_m^{-1}R_{m-1}(\mu_m),&& Z_m^+\equiv\ein_k-2
\Sigma_m^{-1}R_m(\mu_m),\label{eq:Zedef}\\R_m(z)\equiv\half(R_m^+(z)+
R_m^-(z)),&&\Sigma_m\equiv R_m^-(\mu_m)+R_{m-1}^+(\mu_m)+S_m.\nonumber
\eeqa
This result is valid up to {\em exponential} corrections as long as $\vec x$ 
is far removed from all constituents of type $m'\neq m$. Note that in this 
limit $Z_m^{\pm}=\ein_k$, however, only up to {\em algebraic} corrections, 
which is why we have kept them. In Fig.~\ref{fig:zmlocal} we give the 
zero-mode densities for a charge 2 axially symmetric solution in $SU(2)$, 
with well-separated constituents. We found no differences to 1 part in 
$10^6$ between the exact result and the result obtained with the zero-mode 
limit, \refeq{zmlim}. The choice of basis for the zero-modes involves the 
gauge rotation that for each interval $z\in[\mu_m,\mu_{m+1}]$ identifies 
the constituent locations, cmp. Sect.~\ref{sec:bipole} and Ref.~\cite{BrvB}. 
This ensures that each zero-mode is localized to only one constituent monopole. 

It is now almost trivial to read off from this the far field limit for 
$f_x(z,z)$, where $\vec x$ is assumed to be far from {\em all} constituents. 
As long as $z\neq\mu_m$ and $z\neq\mu_{m+1}$, $e_m^-(z)$ and $\tilde e_m^-(z)$ 
are exponentially small, and $f^\ff_x(z,z)=\pi e_m^+(z)(e_m^+)^{-1}\tilde 
e_m^+(z)R_m^{-1}(z)$. But using the definitions of $e_m^\pm(z)$ and 
$\tilde e_m^\pm(z)$ all exponential factors precisely cancel and in 
the far field limit we are left with
\beq
f^\ff_x(z,z)=\pi R_m^{-1}(z),\quad z\in(\mu_m,\mu_{m+1}),\label{eq:fzzffl}
\eeq
which will play a very important role in Sects.~\ref{sec:cons} and 
\ref{sec:chtwo}. We can also read off the far field limit for $f_x(z,z')$ 
evaluated at the impurities $\mu_m$ and $\mu_{m+1}$, noting that by 
definition $e_m^\pm(\mu_{m+1})=\tilde e_m^\pm(\mu_m)=\ein_k$,
\beqa
f^\ff_x(\mu_{m+1},\mu_{m+1})&=&\pi(\ein_k-Z_{m+1}^-)R_m^{-1}(\mu_{m+1})=2\pi
\Sigma_{m+1}^{-1},\nonumber\\f^\ff_x(\mu_m,\mu_m)&=&\pi(\ein_k-Z_m^+)
R_m^{-1}(\mu_m)=2\pi\Sigma_m^{-1},\label{eq:fmmffl}
\eeqa
and $f^\ff_x(\mu_{m+1},\mu_m)=0$ (as well as $f^\ff_x(\mu_m,\mu_{m+1})=0$,
using $f_x(z,z')=f_x^\dagger(z',z)$), verifying the results of 
Ref.~\cite{BrvB}. Although $f_x(z,z)$ is continuous at the impurities, 
comparing \refeq{fzzffl} with \refeq{fmmffl} we see, as anticipated, that 
in the high temperature limit $f_x(z,z)$ is discontinuous at the impurities. 
At finite temperature, the transition across the impurity has a ``width" 
inversely proportional to the temperature.

For $k=1$, where $e_m^\pm(z)=\exp(\pm 2\pi(\mu_{m+1}-z)r_m)$ and 
$\tilde e_m^\pm(z)=\exp(\pm 2\pi(z-\mu_m)r_m)$ one finds for 
$z',z\in(\mu_m,\mu_{m+1})$
\beq
f^\zm_x(z',z)=\frac{2\pi\sinh\left(2\pi r_m(\mu_{m+1}-z'+\gamma^-_{m+1})\right)
\sinh\left(2\pi r_m(z-\mu_m+\gamma^+_m)\right)}{r_m\sinh\left(2\pi r_m
(\nu_m+\gamma_m^++\gamma_{m+1}^-)\right)},
\eeq
which agrees with the result of Ref.~\cite{MTP}, where only the limit with
$\gamma_m^+\equiv-\half\log Z_m^+$ and $\gamma_{m+1}^-\equiv-\half\log
Z_{m+1}^-$ neglected and $z'=z$ was considered. We stress again that the 
presence of $\gamma$ implies a subtle algebraically decaying influence 
due to the constituents of type $m-1$ and $m+1$ (the influence of all other 
constituents is decaying exponentially, although this is only relevant for 
$SU(n>3)$). In terms of constituent radii $r_m=|\vec x-\vec y_m|$ and 
locations $\vec y_m$ one has 
\beq
Z_{m+1}^-=\frac{|\vec y_{m+1}-\vec y_m|+r_{m+1}-r_m}{|\vec y_{m+1}-
\vec y_m|+r_{m+1}+r_m},\quad Z_m^+=\frac{|\vec y_{m-1}-\vec y_m|+
r_{m-1}-r_m}{|\vec y_{m-1}-\vec y_m|+r_{m-1}+r_m}.
\eeq
For $SU(2)$, with $\vec y_{m+1}=\vec y_{m-1}$, therefore $Z_m^+=Z_{m+1}^-$, 
or $\gamma_m^+=\gamma_{m+1}^-$, and the influence of the other constituent 
is only felt by a renormalization of the mass $\nu_m$, 
\beq
SU(2):\  f^\zm_x(z,z)=\pi\frac{\cosh\left(2\pi r_m(\nu_m+2\gamma_m)\right)-
\cosh\left(2\pi r_m(2z-\mu_m-\mu_{m+1})\right)}{r_m\sinh\left(2\pi r_m(\nu_m+
2\gamma_m)\right)}.
\eeq
Note that for $SU(2)$ $\mu_1+\mu_2=0$ and $\mu_2+\mu_3=1$. This leads to the 
well-know result for the monopole zero-mode density $f^\zm_x(z,z)=\pi 
r^{-1}_m\tanh\left(\pi r_m(\nu_m+2\gamma_m)\right)$, with $z=0$ for $m=1$ 
(periodic zero-mode) and $z=1/2$ for $m=2$ (anti-periodic zero-mode).

\section{Bipole zero-modes}\label{sec:bipole}

In this section we discuss the zero-modes in the high temperature limit for
$z=\mu_m$, which means that $\pl e^{-2\pi iz}$ has one of its eigenvalues
equal to 1, which leads to one of the components of the fermion to become
massless. Indeed, using Eqs.~(\ref{eq:zdens},\ref{eq:dftrans},\ref{eq:fmmffl}) 
we find that in the far field limit
\beq
z=\mu_m:\qquad\hat\Psi_z^a(x)^\dagger\hat\Psi_z^b(x)=-(2\pi)^{-1}
\partial_i^2\left(\hat\Sigma_m^{-1}\right)^{ab},\quad\hat\Sigma_m\equiv
\hat g^\dagger(\mu_m)\Sigma_m\hat g(\mu_m),\label{eq:dens}
\eeq
decays algebraically and has support on the constituents of type $m-1$ and 
$m$, as is easy to see for $k=1$, where $\Sigma_m=|\vec x-\vec y_{m-1}|+
|\vec x-\vec y_m|+|\vec y_{m-1}-\vec y_m|$. Here we will restrict ourselves 
to $SU(2)$, particularly interesting for the axially symmetric caloron 
solutions, since in the high temperature limit its gauge field has the form 
of the so-called bipole ansatz (see \refeq{bipans}), which always has an 
integrable chiral fermion zero-mode~\cite{Bip}. In the bipole ansatz all Dirac 
strings have to run in the same direction, but other than that, the locations 
of the self-dual Dirac monopoles can be arbitrary. However, for the axially 
symmetric caloron solutions the constituents have to alternate between 
opposite charges on a line~\cite{BrvB}. In this case, with the solution coming 
from a regular caloron, there are always as many zero-modes as the number of
constituents with a given charge (equal to the topological charge $k$). 
By considering the case of solutions with topological charge 2, we will 
find the expressions for the bipole zero-mode and the extra zero-mode, in 
terms of the constituent locations only (which should be possible, since 
the abelian gauge field has this property in the high temperature limit). 
Remarkably, we will find that rearranging the order of the constituents, 
so as to violate the constraint coming from the axially symmetric caloron, 
the second zero-mode is no longer integrable (while the gauge field and 
bipole zero-mode remain well defined).

A particular class of axially symmetric caloron solutions is obtained by 
taking $Y^j_m\equiv\hat g(z)\frac{\hat A_j(z)}{2\pi i}\hat g^\dagger(z)$ 
to be piecewize constant. This can be shown to satisfy the Nahm or ADHM 
equation when we take~\cite{BrvB}
\beq
\zeta_a=\rho_a\exp(2\pi i\alpha_a)\zeta,\quad\alpha_a\equiv\sum_{m=1}^n
\alpha_a^m P_m,\quad\Tr_n\alpha_a=0,\quad \vec Y_m=Y_m\vec e,\label{eq:zadef}
\eeq
where $\rho_a$ are positive, not to be confused with 
\beq
\vec\rho_m^{\,ab}=\rho^a\rho^b\exp(2\pi i(\alpha_b^m-\alpha_a^m))\Delta
\vec y_m,\quad\Delta\vec y_m=\Delta y_m\vec e\equiv-\pi\tr_2(\zeta^\dagger 
P_m\zeta\vec\tau).\label{eq:rhodef}
\eeq
For $SU(2)$ one has $\Delta\vec y_1=-\Delta\vec y_2$ (for $SU(n>2)$ a
constraint on $\zeta$ is required, to guarantee that all $\Delta\vec y_m$ 
are parallel). We will take $\vec e=\Delta\vec y_2/|\Delta\vec y_2|=(0,0,1)$
(hence $\Delta y_2>0$) and $\pl=\exp(2\pi i\mu_2\tau_3)$ (therefore $\zeta
=\sqrt{\Delta y_2/\pi}\ein_2$). This can always be arranged to be the case 
by a global gauge rotation, and a spatial rotation. We define $\vec y_m$, 
up to an irrelevant overall shift, through $\Delta\vec y_m\equiv\vec y_m-
\vec y_{m-1}$. This fixes $Y_m$ to be
\beq
Y_m^{ab}=(\xi_a+\rho_a^2 y_m)\delta_{ab}+i(1-\delta_{ab})\rho_a\rho_b
\sum_{j= 1}^n\Delta y_j\frac{\exp\left(2\pi i\left[\alpha_b^j-\alpha_a^j
-(\mu_j+s_j^m)(\xi_0^b-\xi_0^a)\right]\right)}{2\sin\left(\pi
\left[\xi_0^b-\xi_0^a\right]\right)}.\label{eq:Ymfull}
\eeq
The constituent locations are found from the eigenvalues of $Y_m$. Although 
constant, it is not true in general that the $Y_m$ can be diagonalized 
simultaneously, making this a non-abelian solution of the Nahm equation. 
Yet, as we remarked before, one easily computes the Green's function, 
since $R^\pm_m(z)=R_m=\sqrt{(\vec x\ein_k-\vec e\,Y_m)\cdot(\vec x\ein_k-
\vec e\,Y_m)}$ is constant in $z$. Using that for the axially symmetric 
solutions $\vec\rho_m^{\,ab}=\hat S_m^{ab}\vec y_m/|\Delta\vec y_m|$, one 
shows~\cite{BrvB} that in the high temperature limit the (abelian) gauge 
field can be written in the form of the bipole ansatz~\cite{Bip}
\beq
A_\mu(x)=-\frac{i}{2}\tau_3\bar\eta^3_{\mu\nu}\partial_\nu\log\phi(x),
\quad\phi(x)=\phi_\ff(x)\equiv\frac{\det(R_1+R_2+S_2)}{\det(R_1+R_2-S_2)}.
\label{eq:bipans}
\eeq
In the bipole ansatz, one splits $A_\mu(x)$ in an isospin up and isospin down 
component (with inverted abelian charges). However, all that concerns us here 
is the fact that for {\em any} $\phi$, $A_\mu(x)$ as given above is self-dual 
(and hence a solution of the Maxwell equations) provided $\log\phi$ is 
harmonic away from Dirac string singularities (defined by $\phi^{-1}=0$). 
This always gives rise to at least one normalizable zero-mode of the chiral 
Dirac equation
\beq
\hat\Psi_{mI}(x)=(2\pi\tilde\rho)^{-1}\phi^{-\hhalf}(x)(\tau_1\bar\sigma_\mu
\veps)_{mI}\partial_\mu\log\phi(x).\label{eq:bipzm}
\eeq
were $\tilde\rho$ is simply a normalization factor. Here $m$ corresponds to the 
isospin component that survives for $z=\mu_m$ (with the other component
related to $\hat f_x(\mu_1,\mu_2)$ vanishing in the far field limit,
see \refeq{Psihat}). 

We now work out the explicit form of all $k$ zero-modes for the axially 
symmetric caloron solution, showing how the zero-mode in \refeq{bipzm} is
recovered from these. Using Eqs.~(\ref{eq:zm},\ref{eq:fmmffl},\ref{eq:rhodef}), 
together with the fact that $\hat f_x(\mu_1,\mu_2)$ is exponentially small
($\tau_1$ is used for picking out the surviving component), we find for the 
normalized zero-modes at $z=\mu_m$ ($|\zeta|=\sqrt{\Delta y_2/\pi}$)
\beq
\hat\Psi_{mI}^a(x)=\phi_\ff^{\hhalf}(x)|\zeta|\rho_b e^{2\pi i\alpha_b^m}
(\tau_1\bar\sigma_\mu\veps)_{mI}\partial_\mu\left(\hat\Sigma_m^{-1}
\right)^{ba}.\label{eq:axzm}
\eeq
Using the fact that (cmp. \refeq{defA})
\beq
\phi^{-1}_\ff(x)=1-2\pi|\zeta|^2\rho_a e^{2\pi i\alpha_a^m}\left(
\hat\Sigma_m^{-1}\right)^{ab}e^{-2\pi i\alpha_b^m}\rho_b,
\eeq
the following linear combination recovers the bipole zero-mode
\beq
\hat\Psi_{mI}(x)=\frac{\rho_a e^{-2\pi i\alpha_a^m}}{\tilde\rho}\hat
\Psi_{mI}^a(x)=\frac{\phi_\ff^{\hhalf}(x)}{2\pi\tilde\rho}(\tau_1\bar
\sigma_\mu\veps)_{mI}\partial_\mu\left(1-\phi_\ff^{-1}(x)\right),\quad
\tilde\rho^2\equiv\sum_a\rho_a^2|\zeta|^2.
\eeq
By defining $\hat\Psi^{(j)}_{mI}(x)=\rho^{(j)}_a\hat\Psi_{mI}^a(x)$ with 
$\rho^{(j)}$ an orthonormal set of complex vectors, with $\rho^{(1)}_a\equiv
\rho_a e^{2\pi i\alpha_a^m}|\zeta|/\tilde\rho$ we form a complete set of 
orthonormal zero-modes. That these are solutions of \refeq{Dirac} is guaranteed
by the general formalism we developed, but one may of course check this by 
substitution in the Dirac equation. This requires $\partial_i\det\Phi
\partial_i\Phi^{-1}=0$, with $\Phi\equiv\left(\ein_k-2\hat S_m
\hat\Sigma_m^{-1}\right)^{-1}$.

\begin{figure}[htb]
\vskip6.7cm
\includegraphics{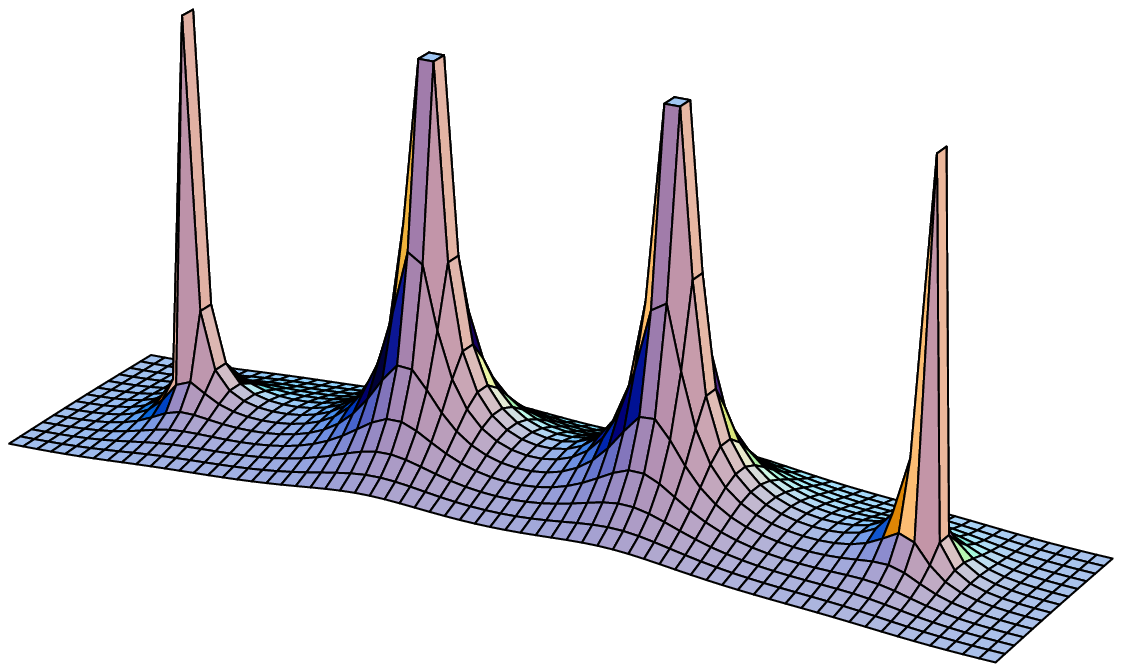}
\includegraphics{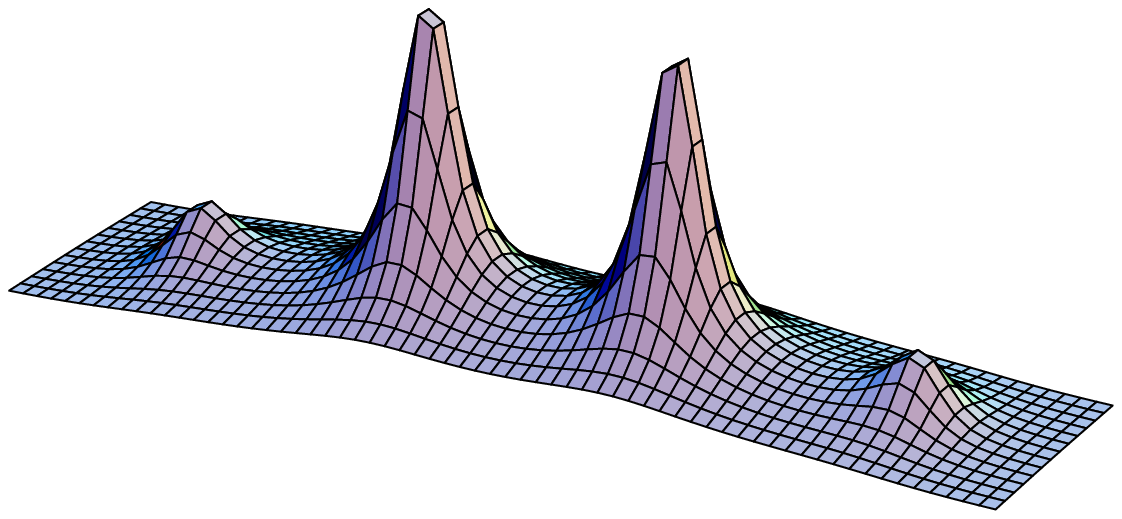}
\includegraphics{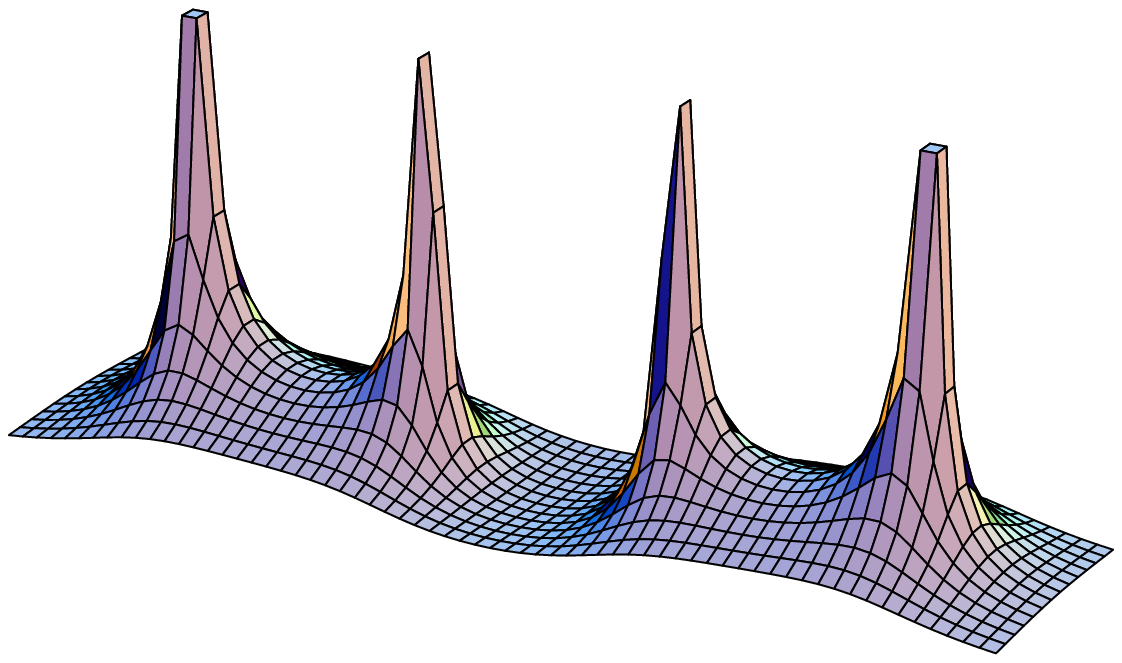}
\includegraphics{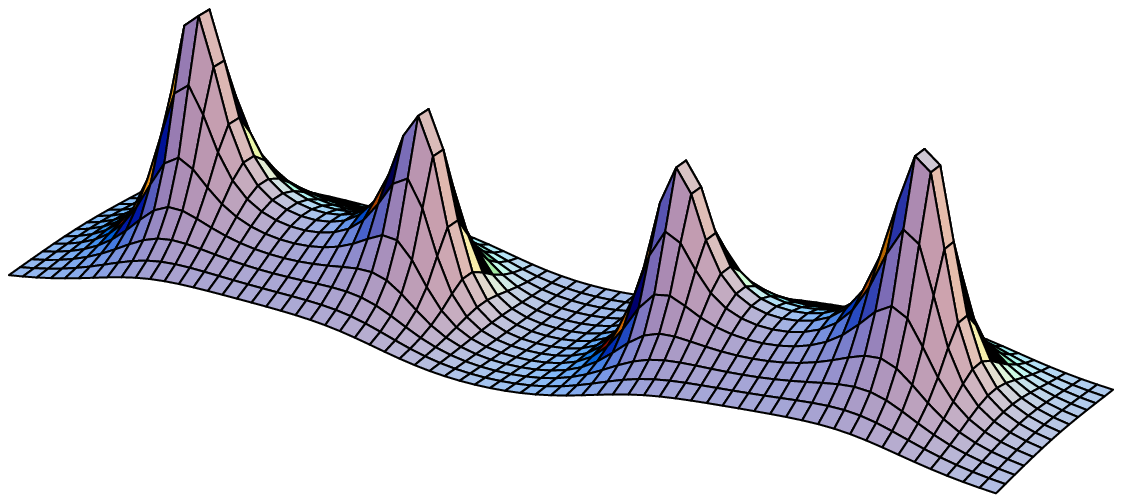}
\caption{The two zero-mode densities at $z=\mu_2=1/4$  (same configuration
as Fig.~\ref{fig:zmlocal}). The bipole zero-mode (right) is at 5 times the 
vertical scale of the second zero-mode (left). Top for the high temperature 
limit and bottom for finite temperature ($\beta=1$).}\label{fig:zmdeloc}
\vspace{4.7cm}
\includegraphics{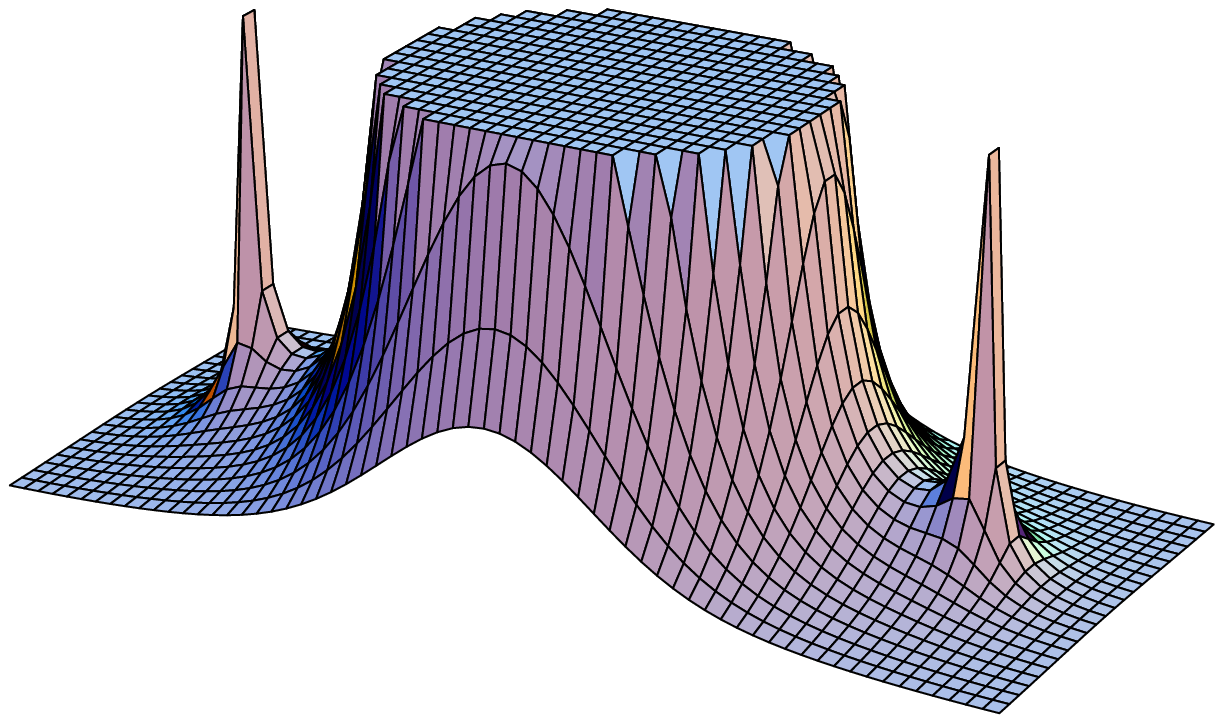}
\includegraphics{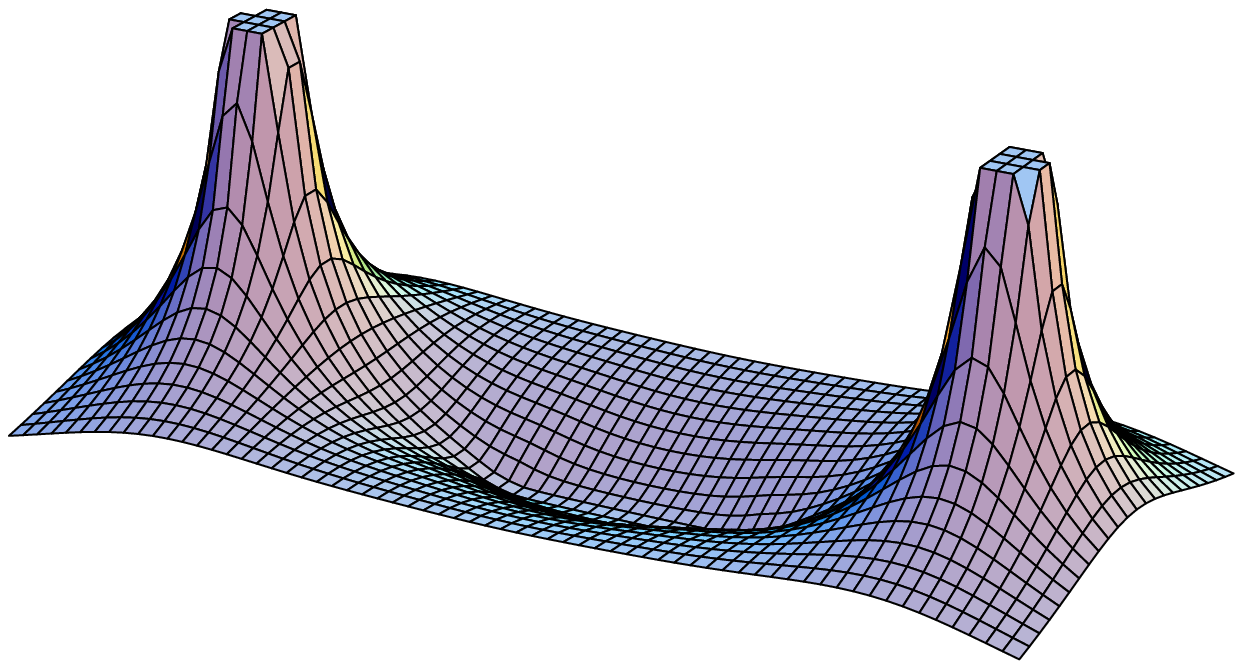}
\caption{The two zero-mode densities for \refeq{phis} with the order of 
$y_1^{(2)}$ and $y_2^{(1)}$ interchanged. The second zero-mode (left) is 
at the same scale, whereas the bipole zero-mode (right) has its 
vertical scale magnified by a factor 10, compared to Fig.~\ref{fig:zmdeloc}.
}\label{fig:singzm}
\end{figure}

For axially symmetric $SU(2)$ solutions with charge 2 we choose $\rho^{(2)}_a
=\veps_{ab}\rho^{(1)}_b$ and find for the two orthonormal zero-modes at 
$z=\mu_2$
\beq
\hat\Psi^{(i)}=\frac{1}{2\pi\tilde\rho}\phi_\ff^\hhalf(x)\pmatrix{\partial_2+
i\partial_1\cr-i\partial_3\cr}\phi_{(i)}^{-1}(x),\quad\phi_{(i)}^{-1}(x)\equiv
2\pi\tilde\rho^2\left(\rho^{(i)}\right)^\dagger\hat\Sigma_2^{-1}\rho^{(1)}.
\eeq
with $\phi_{(1)}^{-1}=1-\phi_\ff^{-1}$, as shown in Ref.~\cite{BrvB}, only 
depending on the constituent locations $y_m^{(a)}$ read off from the 
eigenvalues of $Y_m$. Many choices of $y_m$, $\xi_a$, $\xi_0^a$, $\alpha_a^2$, 
$\rho_a$ and $\mu_2$ actually give rise to the {\em same} constituent 
locations, and hence the {\em same} expressions for $A_\mu(x)$ and 
$\hat\Psi^{(1)}$. It is important for consistency that this will hold for 
$\hat\Psi^{(2)}$ as well. Apart from an irrelevant phase, this is indeed the 
case (checked for many random choices of the parameters). The explicit analytic 
formulae in terms of the 4 arbitrary constituent locations, apart from the 
constraint on the ordering $y_1^{(1)}<y_2^{(1)}<y_1^{(2)}<y_2^{(2)}$, read
\beqa
\phi_{(1)}^{-1}(x)&=&1-\phi_\ff^{-1}=1-\prod_{i=1}^2\frac{r_1^{(i)}+r_2^{(i)}
-|y_1^{(i)}-y_2^{(i)}|}{r_1^{(i)}+r_2^{(i)}+|y_1^{(i)}-y_2^{(i)}|},
\label{eq:phis}\\ \phi_{(2)}^{-1}(x)&=&e^{i\gamma}N\frac{(y_1^{(2)}-y_1^{(1)})
(r_2^{(1)}-r_2^{(2)})+(y_2^{(2)}-y_2^{(1)})(r_1^{(1)}-r_1^{(2)})}{
\sum_{i,j=1}^2N_i^{(j)}r_i^{(j)}}\phi_{(1)}^{-1},\nonumber
\eeqa
where $r_i^{(j)}=|\vec x-\vec e y_i^{(j)}|$ and the constants $N,N_i^{(j)}$ 
are given by
\beq
N\equiv\sqrt{\frac{(y_2^{(2)}-y_1^{(1)})(y_1^{(2)}-y_2^{(1)})
                 }{(y_2^{(1)}-y_1^{(1)})(y_2^{(2)}-y_1^{(2)})}},\ 
N_i^{(j)}\equiv\frac{(y_{i'}^{(j)}-y_{i'}^{(j')})(y_{i'}^{(j)}-y_{i}^{(j')})}{
                     (y_{2}^{(j)}-y_{1}^{(j)})},\ j'\neq j,i'\neq i.
\eeq
The phase $\gamma$ vanishes when $\alpha_a^m=\xi_0^a=0$, but is of course 
irrelevant for checking $\hat\Psi^{(2)}$ to be a properly normalized fermion
zero-mode, orthogonal to $\hat\Psi^{(1)}$. 
In Fig.~\ref{fig:zmdeloc} we give an example for the behavior of these 
zero-mode densities. We choose $y_1^{(1)}=-6.031$, $y_2^{(1)}=-2.031$, 
$y_1^{(2)}=2.031$ and $y_2^{(2)}=6.031$. These are the constituent locations 
also found in Fig.~\ref{fig:zmlocal}, based on the axially symmetric solution 
with $\mu_2=1/4$, $\alpha_a=\xi_0=0$, $\xi=3.5$, $\Delta y_2=1$ and $\rho_1
=\rho_2=2$. Shown are the results for both zero-modes (bipole zero-mode on the 
right) at finite temperature, $\beta=1$ (bottom), and at infinite temperature 
(top). Note that these two only differ in the cores of the constituents, 
regular at finite temperature, but singular for the self-dual Dirac monopoles 
one is left with in the high temperature limit. The bipole zero-mode density 
is shown on a scale enhanced by a factor 5. Its reduced height is due to 
the fact that this zero-mode decays much slower than the other one, as 
can be read off from the behavior of $\phi_{(i)}^{-1}(x)$ in \refeq{phis}.

Crucial for the normalizability of both zero-modes is that $1/\phi_{(i)}$ is 
constant on the Dirac strings, where $1/\phi_\ff$ vanishes 
\beqa
\phi_{(1)}^{-1}(0,0,x_3)=1,&\phi_{(2)}^{-1}(0,0,x_3)=-e^{i\gamma}N\left(
\frac{y_2^{(2)}-y_1^{(2)}}{y_2^{(2)}-y_1^{(1)}}\right)\quad\mbox{for}\quad 
y_1^{(1)}\leq x_3\leq y_2^{(1)},\nonumber\\\phi_{(1)}^{-1}(0,0,x_3)=1,&
\phi_{(2)}^{-1}(0,0,x_3)=~~e^{i\gamma}N\left(\frac{y_2^{(1)}-y_1^{(1)}}{y_2^{
(2)}-y_1^{(1)}}\right)\quad\mbox{for}\quad y_1^{(2)}\leq x_3\leq y_2^{(2)}.
\eeqa
We are now in a position to answer the question what happens when violating 
the constraint on the alternating order of oppositely charged constituents,
by smoothly deforming from $y_2^{(1)}<y_1^{(2)}$ to $y_2^{(1)}>y_1^{(2)}$.
Under this deformation, both $\hat\Psi^{(1)}$ and $\hat\Psi^{(2)}$ remain 
zero-mode solutions of the Dirac equation, and $A_\mu$ remains a self-dual 
abelian gauge field. We now have a Dirac string for $y_1^{(2)}<x_3<y_2^{(1)}$ 
of double the usual strength ($\phi_\ff^{-1}$ behaving as $x_\perp^4$ as 
opposed to $x_\perp^2$), where the second zero-mode density diverges, as 
illustrated in Fig.~\ref{fig:singzm}. This is because $\phi_{(2)}$ will no 
longer be constant on the double Dirac string. The bipole zero-mode, on the 
other hand, remains well defined. It actually vanishes identically on the 
double Dirac string (cmp. Fig.~1 of Ref.~\cite{Bip}), and no longer ``sees" 
the two inner self-dual Dirac monopoles.

It would be interesting if one could formulate an index theorem for these 
abelian field configurations with singularities, but this will not be 
straightforward as our analysis shows. It is yet another subtlety in 
describing the monopole content of non-abelian gauge fields. Developing 
a better understanding of these constraints, that affect the long range 
properties of configurations, is our main motivation for these studies. 

\section{Appearance of conserved quantities}\label{sec:cons}

Our analysis has shown that in all cases, as long as $z$ stays away from the 
impurities, the zero-mode density is exponentially localized to the cores of 
the constituents in the far field limit. The sizes of the monopole cores 
shrink to zero in the high temperature limit (with masses scaling proportional 
with the temperature), therefore we expect 
\beq
\cV_m(\vec x)\equiv(4\pi)^{-1}\Tr\left(R_m^{-1}(z)\right),
\label{eq:Rcons}
\eeq
cmp. Eqs.~\ref{eq:zdens} and \ref{eq:fzzffl}, to be harmonic almost everywhere
except for singularities tracing the cores of type $m$ monopoles for $z\in
( \mu_m,\mu_{m+1})$. Since the caloron gauge field does not depend on $z$, 
this interpretation requires $\cV_m$ not to depend on $z$. The trace in 
\refeq{Rcons} is necessary to remove any $z$ dependence due to the fact that 
the basis of zero-modes is only defined up to a -- possibly $z$ dependent --
unitary transformation. All this is obviously true when $\vec R(z;\vec x)$ 
is piecewize constant, as for $k=1$ and for the class of axially symmetric 
solutions of Ref.~\cite{BrvB}. In this case the zero-mode density in the 
high temperature limit reduces to the sum of $k$ delta function, located 
at the appropriate constituent monopole locations.

To show that $\cV_m$ is independent of $z$, even when $\vec R(z,\vec x)$ is 
{\em not} piecewize constant we solve the Riccati equation, \refeq{Riccati},
iteratively in $1/|\vec x|$ and obtain the multipole expansion for $\cV_m
(\vec x)$. We can then use the Nahm equation for $\hat A_j(z)$ to check if 
every moment of this expansion is independent of $z$. We restrict our 
attention to the region between $\mu_m$ and $\mu_{m+1}$ and perform a 
rotation $U(\hat x)$ 
\beq
Y_i(z)\equiv U_{ij}(\hat x)\hat g(z)\frac{\hat A_j(z)}{2\pi i}\hat 
g^\dagger(z),\quad
U(\hat x)=\pmatrix{\sin(\theta)\cos(\vphi)&\sin(\theta)\sin(\vphi)&\hphantom{-}
  \cos(\theta)\cr\cos(\theta)\cos(\vphi)&\cos(\theta)\sin(\vphi)&-
  \sin(\theta)\cr-\sin(\vphi)&\cos(\vphi)&0\cr},\label{eq:defU}
\eeq
with $(\theta,\vphi)$ defined such that $U_{1j}=\hat x_j\equiv x_j/|\vec x|$
(the dependence of $\vec Y(z)$ on $\hat x$ will always be implicitly assumed).
This leaves rotations around $\hat x$ as a remaining freedom, which we will
make use of later. For the axially symmetric solutions discussed in 
Sect.~\ref{sec:bipole}, $\vec Y(z)=U(\hat x)\vec Y_m$, cmp. \refeq{Ymfull}. 

The Nahm equation, which is invariant under rotations, 
is equivalent to (working in the gauge where $\hat A_0$ is constant, removed 
by the gauge transformation with $\hat g(z)$)
\beq
\ddz Y_i(z)=-\pi i\veps_{ijk}[Y_j(z),Y_k(z)].\label{eq:snahm}
\eeq
We introduced $\vec Y(z)$ in \refeq{defU} such that 
\beq
\vec R(z;\vec x)^2=\ein_k|\vec x|^2-2|\vec x|Y_1(z)+\vec Y^{\,2}(z),
\eeq
has a simple form. Writing
\beq
R_m^\pm(z)^2\equiv\ein_k|\vec x|^2-|\vec x|Q_\pm(z;|\vec x|^{-1}),
\label{eq:wdef}
\eeq
we can now formulate the Riccati equation, \refeq{Riccati}, in terms of 
$\vec Y(z)$ and $Q_\pm(z;|\vec x|^{-1})$, 
\beq
\ein_k-\frac{Q_\pm(z;|\vec x|^{-1})}{|\vec x|}=\ein_k-2\frac{Y_1(z)}{|\vec x|}+
\frac{\vec Y^{\,2}(z)}{|\vec x|^2}\pm\frac{1}{2\pi|\vec x|}\ddz\sqrt{\ein_k-
\frac{Q_\pm(z;|\vec x|^{-1})}{|\vec x|}},\label{eq:QRic}
\eeq
which can be solved by iteration, expanding in powers of $1/|\vec x|$, 
something that is easily automated. We used the algebraic program 
FORM~\cite{FORM} for its superior memory management and speed to push
this calculation to a high order. We find for the first few terms (also
easily obtained by hand),
\beqa
R_m^\pm(z)/|\vec x|&=&\ein_k-Y_1/|\vec x|+\half
(Y_2^2+Y_3^2\pm i[Y_3,Y_2])/|\vec x|^2+\\&&\half(Y_2Y_1Y_2+Y_3Y_1Y_3\mp i
Y_2Y_1Y_3\pm iY_3Y_1Y_2)/|\vec x|^3+\ldots\label{eq:Qs}\nonumber
\eeqa
where the $z$ dependence of $Y_j(z)$ is suppressed for ease of notation
and any derivatives with respect to $z$ are eliminated with the help of the
Nahm equation, \refeq{snahm}. We substitute this expansion in \refeq{Rcons},
with $R_m(z)=\half(R_m^+(z)+R_m^-(z))$, from which we obtain its multipole 
expansion. The first few terms are 
\beq
\cV_m(\vec x)=\frac{1}{4\pi|\vec x|}\Tr\left(\ein_k+Y_1/|\vec x|+\half
(3Y_1^2-\vec Y^{\,2})/|\vec x|^2+\half Y_1(5Y_1^2-3\vec Y^{\,2})
/|\vec x|^3+\ldots\right).\label{eq:Rexp}
\eeq

A number of checks can be performed on this result. First of all $\cV_m$ 
and $Q_\pm$ are invariant under any rotation among $Y_2$ and $Y_3$,
$\cV_m$ is real and $Q_\pm$ are hermitian, as should be. By construction,
cmp. Eqs.~(\ref{eq:zdens}), (\ref{eq:fzzffl}) and (\ref{eq:Rcons}),
\beq
\sum_{a=1}^k\hat\Psi_z^a(x)^\dagger\hat\Psi_z^a(x)=-\partial_i^2\cV_m(\vec x),
\quad z\in(\mu_m,\mu_{m+1})\label{eq:Trzdens}
\eeq
in the far field limit, and one verifies that this integrates to $k$, 
as required by the proper normalization of the zero-modes. But most 
importantly, we verify with the help of the Nahm equation, \refeq{snahm}, that 
$\ddz\cV_m=0$ to the order given (whereas in general neither $R^\pm_m(z)$, 
nor $\Tr(R_m(z))$ are constant). For the dipole term this is immediate, 
since $\ddz\Tr Y_1=-2\pi i \Tr[Y_2,Y_3]=0$. For the quadrupole term we have 
$\ddz\Tr(3Y_1^2-\vec Y^{\,2})=-4\pi i\Tr(3Y_1[Y_2,Y_3]-\veps_{ijk}Y_iY_jY_k)
=0$, using the cyclic property of the trace, etc. Note that $\Tr\hat A_i(z)$ 
plays a special role. It corresponds to the $U(1)$ part of the $U(k)$ Nahm 
gauge field, and therefore decouples in the Nahm equations. As is obvious 
from the definition of $\vec R(z;\vec x)$, \refeq{Rdef}, it can actually 
be absorbed in a shift of $\vec x$. Therefore, where this simplifies matters 
we may assume $\vec Y(z)$ to be traceless. 

Finally we check, as conjectured above, that each of the terms is harmonic. 
For this we have to note that $\vec Y(z)$ depends on $\vec x$ through the 
rotation $U(\hat x)$, see \refeq{defU}. In the expression for $\cV_m(\vec x)$ 
the $\vec x$ dependence is easily recovered since $\vec Y^{\,2}$ is independent 
of $\vec x$ and $Y_1(z)=\hat g(z)\hat A_j(z)\hat g^\dagger(z)x_j/(2\pi i
|\vec x|)$. It is now straightforward to verify that each term is harmonic. 
When $k=1$ we may use that $Y_1=\hat x\cdot\vec y_m$ is no longer a matrix, 
and one indeed finds $\cV_m(\vec x)$ in \refeq{Rexp} to be the multipole 
expansion for $(4\pi|\vec x-\vec y_m|)^{-1}$ for that case. Since this is
harmonic (for $\vec x\neq \vec y_m$), each term in the multipole expansion of 
$\cV_m(\vec x)$ has to be harmonic. For arbitrary $k$ and $\vec Y(z)$ we have 
checked these properties to order $|\vec x|^{-14}$ in the multipole expansion. 

From now on we take charge 2. In this case there are 5 independent conserved 
quantities that characterize the solutions of the Nahm equation on a given 
interval $z\in[\mu_m,\mu_{m+1}]$, apart from the 3 translational degrees of 
freedom contained in $\Tr\hat A_i(z)$. They are given by the entries of the 
traceless and symmetric matrix $M$,
\beq
M_{ij}\equiv-\half\left(\Tr(\hat A_i(z)\hat A_j(z))-\third
\delta_{ij}\Tr(\hat A_k(z)\hat A_k(z))\right).\label{eq:defM}
\eeq
One easily checks with the Nahm equation that this is conserved, as for 
the quadrupole term considered above. Although not needed here, it is 
known~\cite{NCal} that for any $k$ a solution to the Nahm equation implies 
$\det(y_j\hat A_j(z)-2\pi iy_jx_j)$ is constant, provided $\vec y\in C^3$, 
with $y_j^2=0$. Indeed, for $k=2$ and $\vec x=\vec 0$, using that 
$\det(y_j\hat A_j(z))=-\half\Tr(y_j\hat A_j(z))^2$, this is equivalent to 
$M$ being constant if and only if $y_j^2=0$. Since $M$ gives all the 
independent invariants, it should be possible to express $\cV_m(\vec x)$ 
in terms of $M$ and $\vec x$. Considerable simplifications occur, because 
we can write 
\beq
\hat A_j(z)\equiv i\cA_{ja}(z)\hat g^\dagger(z)\tau_a\hat g(z)\label{eq:Aia}
\eeq
(absorbing the trace part in a shift of $\vec x$), and use $\tau_a\tau_b=
\delta_{ab}\ein_2+i\veps_{abc}\tau_c$ to reduce the matrix products to scalar 
products, $\cA_{ia}(z)\cA_{ja}(z)=M_{ij}+\third\delta_{ij}\cA^2_{ka}(z)$.
We do indeed find that $\cV_m(\vec x)$ is a function of $M$ and $\vec x$ only, 
\beq
\cV_m(\vec x)=\frac{1}{2\pi|\vec x|}\left(1+\frac{3}{2|\vec x|^2}\hat M_{11}
(\hat x)+\cO(|\vec x|^{-4})\right).
\eeq
where for convenience we introduced 
\beq
\hat M_{ij}(\hat x)=\frac{1}{4\pi^2}\left(U(\hat x)MU^{-1}(\hat x)\right)_{ij}
=\half\Tr\left(Y_i(z)Y_j(z)-\third\delta_{ij}\vec Y^{\,2}(z)\right).
\label{eq:defMhat}
\eeq
We performed the multipole expansion for charge 2 to order $|\vec x|^{-21}$. 
Only the odd orders appear, because $\vec Y(z)$ is assumed to be traceless.
In appendix B we give the term of order 21, and show how from this all lower 
order multipole coefficients can be recovered.

\section{Exact results for charge 2}\label{sec:chtwo}

In this section we will construct $SU(2)$ charge 2 solutions for which $\vec 
R(z;\vec x)$ is {\em not} piecewize constant and analyse the localization 
of the fermion zero-modes in the far field limit. We already saw in 
Sect.~\ref{sec:cons} that on each interval $(\mu_m,\mu_{m+1})$ we have
information on the zero-mode density, \refeq{Trzdens}, in terms of 8
parameters. Three of these are associated with $\Tr\vec Y(z)$, which give the 
center of mass coordinates for the constituents of type $m$. Of the other 5, 
given by the $3\times 3$  traceless symmetric matrix $M$, 3 are associated to
a rotation $\cR$ that diagonalizes $M$, whereas the remaining 2 parameterize 
the eigenvalues of $M$. They will give a scale ($D$) and shape ($\k$) 
parameter, see below.

\subsection{Solutions to the Nahm equation}\label{sec:k0}
Explicit solutions in terms of Jacobi elliptic functions were first considered 
in the context of $SU(2)$ charge 2 monopoles~\cite{NahmM,BrDa}. 
These solutions can be adopted for the calorons provided the appropriate 
boundary conditions, read off from \refeq{nahm}, are implemented. In terms 
of the $3\times 3$ matrix $\cA_{ia}(z)$ defined in \refeq{Aia}, the Nahm 
equation becomes (away from $z=\mu_m$) 
\beq
\frac{1}{2}\ddz\cA^t(z)=\det(\cA(z))\cA^{-1}(z),
\eeq
from which we find
\beq
\frac{1}{4}\ddz\left(\cA(z)\cA^t(z)\right)=\ein_3\det\cA(z)=\ein_3\sqrt{\det
\left(\cA(z)\cA^t(z)\right)}.\label{eq:AA}
\eeq
The traceless part of $\cA(z)\cA^t(z)$ is therefore independent of $z$, once 
again verifying that $M=\cA(z)\cA^t(z)-\third\ein_3\Tr(\cA(z)\cA^t(z))$ is 
constant. Here we are, however, interested in the equation for the trace part 
\beq
\ddz F(z)=4\sqrt{\det\left(F(z)\ein_3+M\right)},\quad
F(z)\equiv\third\Tr\left(\cA(z)\cA^t(z)\right).
\eeq
When we diagonalize $M$ with a suitable rotation $\cR$, $M=\cR\diag(c_1,c_2,
c_3)\cR^t$, fixing $\cR$ such that $c_2\leq c_1\leq c_3$ (note that in 
addition $c_1+c_2+c_3=\Tr M=0$), this can be cast in the form
\beq
\ddz F(z)=4\sqrt{(F(z)+c_1)(F(z)+c_2)(F(z)+c_3)}.
\eeq
Defining $F(z)+c_i=\quart D^2 f_i^2(Dz)$ and introducing $D$ and 
$\k$ to parameterize the $c_i$,
\beq
D^2\equiv\quart(c_3-c_2),\quad \k^2\equiv\frac{c_3-c_1}{c_3-c_2}\leq 1,
\label{eq:Dkdef}
\eeq
shows that the solution can be written in terms of the Jacobi elliptic 
functions\footnote{The Jacobi elliptic functions are defined by $sn_\k(u)=
\sin(\vphi(u))$, $cn_\k(u)=\cos(\vphi(u))$ and $dn_\k(u)=\sqrt{1-\k^2sn^2_\k
(u)}$, with $u=\int_0^{\vphi(u)}d\theta(1-\k^2\sin^2\theta)^{-\hhalf}$. We 
use boldface for $\k$ to avoid confusion with charge $k$. One also encounters 
the notation~\cite{AbSt} $sn(u|m)$ for $sn_\k(u)$, with $m=\k^2$.} 
\beq
f_1(z)=\frac{\k'}{cn_\k(z)},\quad f_2(z)=\frac{\k'sn_\k(z)}{cn_\k(z)},
\quad f_3(z)=\frac{dn_\k(z)}{cn_\k(z)},\quad\k'\equiv\sqrt{1-\k^2}.
\eeq
The overall sign of the functions $f_i(z)$ is chosen such that $df_1(z)/dz=
f_2(z)f_3(z)$, and cyclic, such that in terms of these the most general 
solution of the Nahm equation is given by
\beq
\hat A_j(z;\vec a,\cR,h,\cD,\k)\equiv 2\pi i\hat g^\dagger(z)h^\dagger
\Bigl(a_j\ein_2+\cD\cR_{jb}f_{b}(Dz)\tau_b\Bigr)h\hat g(z),
\quad\cD\equiv(4\pi)^{-1}D,\label{eq:fnahm}
\eeq
where $\cR$ is the rotation that diagonalizes $M$ and $h$ is a global gauge 
rotation (leaving $\cA(z)\cA^t(z)$ invariant). In the $Sp(1)$ formalism for 
constructing $SU(2)$ calorons one requires $\hat A^t_j(z)=\hat A_j(-z)$, a 
property shared by $f_b(Dz)\tau_b$ for each $b$. Arranging $f_2(z)$ to be 
odd and $f_{1,3}(z)$ to be even in $z$ was the reason for choosing 
$c_2\leq c_1\leq c_3$. 

We see that $\k'=0$ (i.e. $\k=1$) recovers the case where $\vec R(z;\vec x)$ 
is piecewize constant, for which $\cV^{\k=1}_m(\vec x)=(4\pi|\vec x-\vec y|
)^{-1}+(4\pi|\vec x+\vec y|)^{-1}$, with $\vec y=(0,0,\cD)$ and $\pm\vec y$ 
the two locations of the equal charge constituents (the center of mass assumed 
to be zero), see Sect.~\ref{sec:cons}. The combined zero-mode density in the 
far field limit is given by $-\partial_i^2\cV_m(\vec x)=\delta(\vec x-\vec y)
+\delta(\vec x+\vec y)$, the sum of two delta-functions at these constituent 
locations. Actually, two point-like constituents necessarily implies $\k=1$. 
Comparing $3\hat M_{11}/(16\pi^3|\vec x|^3)=3\sum_{j=1}^3c_jx_j^2/(16\pi^3|
\vec x|^5)$, see \refeq{defMhat}, with the quadrupole term for $\cV^{\k=1}_m
(\vec x)$, $(2x_3^2-x_2^2-x_1^2)\cD^2/(4\pi|\vec x|^5)$, one finds that 
$c_1=c_2$, which forces $\k=1$. It is in this context that we define 
$\vec y=\pm(0,0,\cD)$ as would-be constituent locations even when $\k\neq 1$.
In which way we will approach point-like constituents when $\k\to1$ will 
become clear in Sect.~\ref{sec:ext}. So far we have studied the Nahm equation 
away from the ``impurities". We first want to convince ourselves that there 
are more general (not piecewize constant) solutions to the Nahm equation 
that describe calorons with non-trivial holonomy, for which we need to solve 
the boundary conditions at the ``impurities".

\subsection{Matching at the impurities}
We will consider here the case of $SU(2)$ charge 2 calorons in the symplectic 
representation, for which $\hat A_j(-z)=\hat A^t_j(z)$. This condition is 
preserved under the gauge transformation $\hat g(z)$ defined in 
\refeq{dftrans}, but requires us to further constrain $h$ appearing in 
\refeq{fnahm} to be generated by $\tau_2$. It means that for $z\in[\mu_1,
\mu_2]$ the Nahm gauge field is given by $\hat A_j(z)=\hat A_j(z;\vec a^{(1)},
\cR^{(1)},e^{-\frac{i}{2}\theta^{(1)}\tau_2}, \cD^{(1)},\k^{(1)})$. For 
$z\in[\mu_2,1+\mu_1]$ we use that periodicity of $\hat A_j(z)$ implies 
$\hat A_j(1-z)=\hat A_j(-z)=\hat A_j^t(z)$, such that $\hat A_j(z)=
\hat A_j(z-\half;\vec a^{(2)},\cR^{(2)},e^{-\frac{i}{2}\theta^{(2)}
\tau_2},\cD^{(2)},\k^{(2)})$. This was studied before by Houghton and Kraan 
for trivial holonomy ($\mu_2=0$)~\cite{HoKr}, where one of the monopole types 
is massless. For general holonomy, the Nahm equation reduces at $z=\mu_2$ to 
(see \refeq{nahm})
\beq
\hat A_j(\mu_2-\half;\vec a^{(2)},\cR^{(2)},e^{-\frac{i}{2}\theta^{(2)}\tau_2},
\cD^{(2)},\k^{(2)})-\hat A_j(\mu_2;\vec a^{(1)},\cR^{(1)},e^{-\frac{i}{2}
\theta^{(1)}\tau_2},\cD^{(1)},\k^{(1)})=2\pi i\rho_2^j.\label{eq:jump}
\eeq
At $z=\mu_1=-\mu_2$, using $\hat A_j(-z)=\hat A_j(z)^t$, the same condition 
is found (one can check~\cite{BrvB} that $\rho_1=-\rho_2^t$). 

\begin{figure}[htb]
\vspace{4.5cm}
\includegraphics{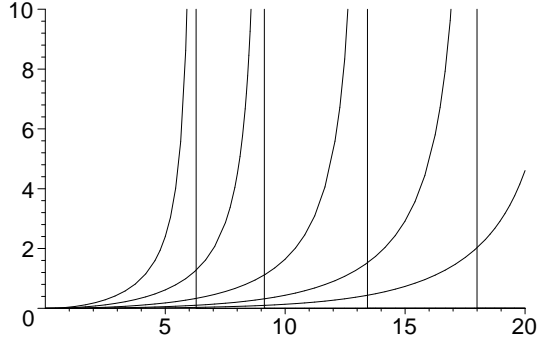}
\caption{Plotting $|\Delta\vec a|$ versus $D$, \refeq{bcI}, for $\k=0,$ 
0.9, 0.99, 0.999 and 0.9999 (left to right).}\label{fig:kDa-plot} 
\end{figure}

A particularly simple solution is obtained for $\mu_2=1/4$ (equal mass 
constituents) by taking the same parameters in both intervals, up to a 
shift $\vec a$, $\cD^{(m)}=\cD$, $\k^{(m)}=\k$, $\cR^{(m)}=\ein_3$ and 
$\theta^{(m)}=0$, collapsing \refeq{jump} to
\beq
\vec\rho_2=\Delta\vec a\ein_2-2(0,1,0)\tau_2\cD\k'\frac{sn_\k(\quart D)}{
cn_\k(\quart D)},\quad\Delta\vec a\equiv\vec a^{(2)}-\vec a^{(1)}
\eeq
This can be solved by taking $P_2=\half(\ein_2+\tau_2)$, $\zeta_a=\rho
\exp(2\pi i\alpha_a\tau_2)$, with $\alpha_1-\alpha_2=1/4$, which
gives $\vec\rho_2^{\,ab}=-2\pi\rho^2P_2^{ab}(0,1,0)$, such that
\beq
\frac{D\k'sn_\k(\quart D)}{2\pi cn_\k(\quart D)}(0,1,0)=
-\Delta\vec a=\pi\rho^2(0,1,0).\label{eq:bcI}
\eeq
For increasing $D$, $cn_\k(\quart D)$ will reach zero at $\quart D=K(\k)$,
the half-period\footnote{$K(\k)=\int_0^{\pi/2}d\theta(1-\k^2\sin^2\theta
)^{-\hhalf}$, satisfying $K(\k)=-\log(\quart\k')(1+\cO(\k'^2))$~\cite{AbSt}}. 
To keep $|\Delta\vec a|$ finite, $\k'$ has to approach zero as well. 

In Fig.~\ref{fig:kDa-plot} we plot $|\Delta\vec a|$ as a function of $\k$ 
and $D$, from which we confirm that, at fixed $|\Delta\vec a|$, $\k$ has 
to approach 1 to have well separated constituents. In this point-like limit, 
monopole constituents of type 1 are located at $(0,-\half\pi \rho^2,\pm\cD)$ 
and those of opposite charge (type 2) at $(0,\half\pi\rho^2,\pm\cD)$ (choosing 
the overall center of mass at the origin). This configuration has parallel 
magnetic moments and placing self-dual Dirac monopoles at the mentioned 
locations would give a gauge field in the bipole ansatz, \refeq{bipans}. 
However, the abelian component of the gauge field coming from the above caloron
can at best take the bipole form in the limit discussed. For $\k=1$ and $\cD$ 
finite, $|\Delta\vec a|=0$, and one is left with two singular instantons (of 
zero size) at $(0,0,\pm\cD)$. 

For this reason we now consider a class of solutions which contains 
{\em regular} axially symmetric solutions with $\k=1$. Again we take 
$\mu_2=1/4$, $\cD^{(m)}=\cD$, $\k^{(m)}=\k$, $\zeta_a=\rho\exp(2\pi 
i\alpha_a\tau_2)$, but now with $\alpha_2=-\alpha_1\equiv\pi^{-1}\alpha$, 
$P_2=\half(1+\tau_3)$, $\theta^{(2)}=-\theta^{(1)}\equiv\theta$ and 
\beq
\cR^{(2)}=\left(\cR^{(1)}\right)^{-1}=\pmatrix{\cos\vphi&0&\sin
\vphi\cr0&1&0\cr-\sin\vphi&0&\cos\vphi}.
\eeq
One finds 
\beq
\vec\rho_2^{\,ab}=-\pi\rho^2(-\sin\alpha\tau_3,-\sin\alpha\tau_2,
\tau_1+\cos\alpha\ein_1)^{ab},
\eeq
and \refeq{jump} takes the following form
\beqa
2\cD(f_1(\quart D)\sin\theta\cos\vphi+f_3(\quart D)\cos\theta\sin\vphi)
\tau_3+\Delta a_1\ein_2&=&\pi\rho^2\sin\alpha\tau_3\nonumber\\
-2\cD f_2(\quart D)\tau_2+\Delta a_2\ein_2&=&\pi\rho^2\sin\alpha\tau_2\\
-2\cD(f_1(\quart D)\cos\theta\sin\vphi+f_3(\quart D)\sin\theta\cos\vphi)
\tau_1+\Delta a_3\ein_2&=&-\pi\rho^2(\tau_1+\cos\alpha\ein_2).\nonumber
\eeqa
This can be simplified to
\beqa
\cD\sin(\theta-\vphi)\Bigl(f_3(\quart D)-f_1(\quart D)\Bigr)=\half\pi\rho^2
(1-\sin\alpha),&\Delta\vec a=-\pi\rho^2\cos\alpha(0,0,1),\label{eq:sol}\\
\cD\sin(\theta+\vphi)\Bigl(f_3(\quart D)+f_1(\quart D)\Bigr)=\half\pi\rho^2
(1+\sin\alpha),&\cD f_2(\quart D)=-\half\pi\rho^2\sin\alpha,\nonumber
\eeqa
It gives a three parameter family of solutions with would-be point-like 
constituents at
\beq
\vec y_m^{\,(j)}=((-1)^{j}\cD\sin\vphi,0,(-1)^{m+j}\cD\cos\vphi-(-1)^m
\half\rho^2\cos\alpha).\label{eq:conloc}
\eeq

To have an exact point-like far field limit we need to impose $\k=1$, 
implying $\sin\alpha=0$ and $\cos\theta\sin\vphi=0$. The first possibility 
is that $\cos\theta=0$, for which $|\cos\vphi|\cD=\half\rho^2$. 
One finds two constituents of opposite charge to coincide. Such a solution 
describes a singular (zero-size) instanton on top of a smooth caloron. 
Excluding this singular case we are left with the choice $\sin\vphi=0$, 
for which $|\sin\theta|\cD=\half\rho^2$. We now find axially symmetric 
solutions with constituent locations at
\beq
\vec y_m^{\,(j)}=\mp\half\pi\rho^2\left((-1)^m+(-1)^j|\sin\theta|^{-1}
\right)(0,0,1), 
\eeq
where the overall sign comes from the fact that $\cos\alpha=\pm1$. For $\cos
\theta\neq0$, all constituents are now separated from each other, giving a 
regular solution. Both cases were already studied in Ref.~\cite{BrvB}, based 
on assuming that $\vec\rho_2$ is one dimensional (cmp. Sect.~\ref{sec:bipole}). 
It can be shown that for $SU(2)$ {\em exact} point-like constituents, 
$\k^{(1)}=\k^{(2)}=1$, forces $\vec\rho_2$ to be one dimensional for 
{\em any} choice of the remaining parameters. 

\begin{figure}[htb]
\vspace{6.5cm}
\includegraphics{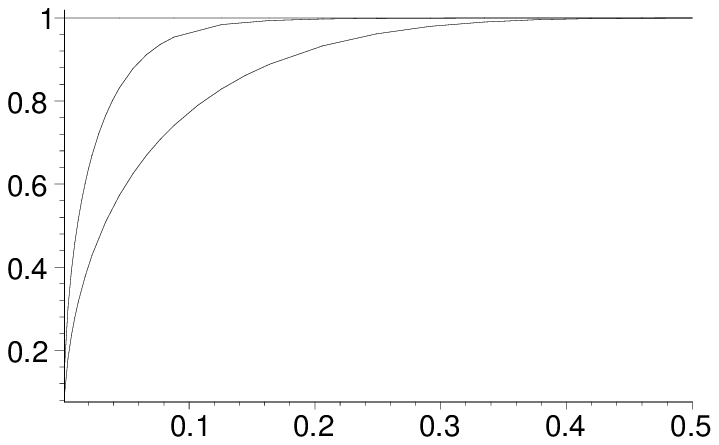}
\includegraphics{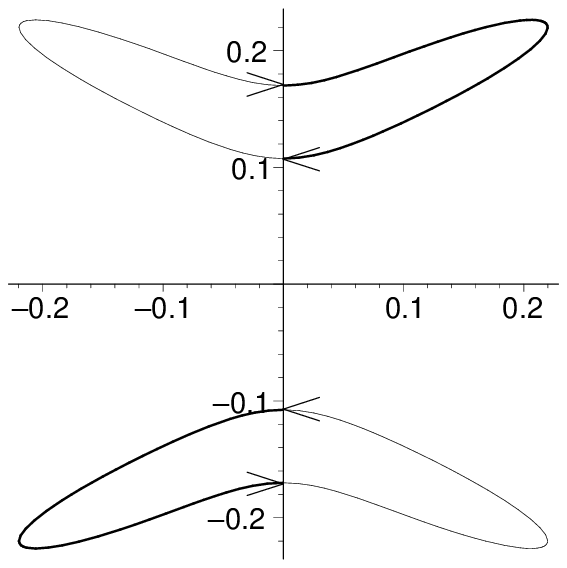}
\caption{On the left we plot $\k$ versus $\rho$ for $\alpha=0$ ($\k\equiv1$), 
$\alpha=-\pi/100$ and $\alpha=-\pi/2$ (giving the lower bound for $\k$ at 
fixed $\rho$). On the right are shown the locations, \refeq{conloc}, of 
monopoles and antimonopoles (fat vs. thin curves) in the $1,3$-plane for 
$\rho=1/4$, by varying $\alpha$ from $-\pi$ (indicated by the arrows) to 0.
}\label{fig:conloc}
\end{figure}

Nevertheless, when $\sin\alpha\neq0$, insisting as before that equal charge 
constituents are well separated, while keeping the centers of mass of these 
pairs at a fixed distance $\pi\rho^2\cos\alpha$, one forces $\k\to1$ while 
increasing $\cD$, and hence approximate point-like constituents. We will 
illustrate this behavior for $\theta=\pi/4$. In Fig.~\ref{fig:conloc} 
we plot for a typical value of $\rho$ the constituent locations as given 
by \refeq{conloc}, varying $\alpha$ between $-\pi$ and 0 (given $\rho$, 
$\alpha$ and $\theta$, one can use \refeq{sol} to solve for $\vphi$, $\cD$ 
and $\k$). We also plot $\k$ as a function of $\rho$ for some values of 
$\vphi$, showing the rapid uniform approach to $\k=1$. The asymptotic 
behavior for $\alpha=-\pi/2$ is determined by 
\beq
\k'=\frac{4\exp(-D/4)}{3+2\sqrt{2}}\left(1+\cO(\k'^2)\right),
\quad D=4\sqrt{2}\pi^2\rho^2\left(1+\cO(\k'^2)\right).\label{eq:kpasym}
\eeq

\begin{figure}[htb]
\vskip11mm\hskip73mm$f_3$\vskip16mm\hskip73mm$f_1$\vskip13mm\hskip73mm$f_2$
\vskip2mm
\includegraphics{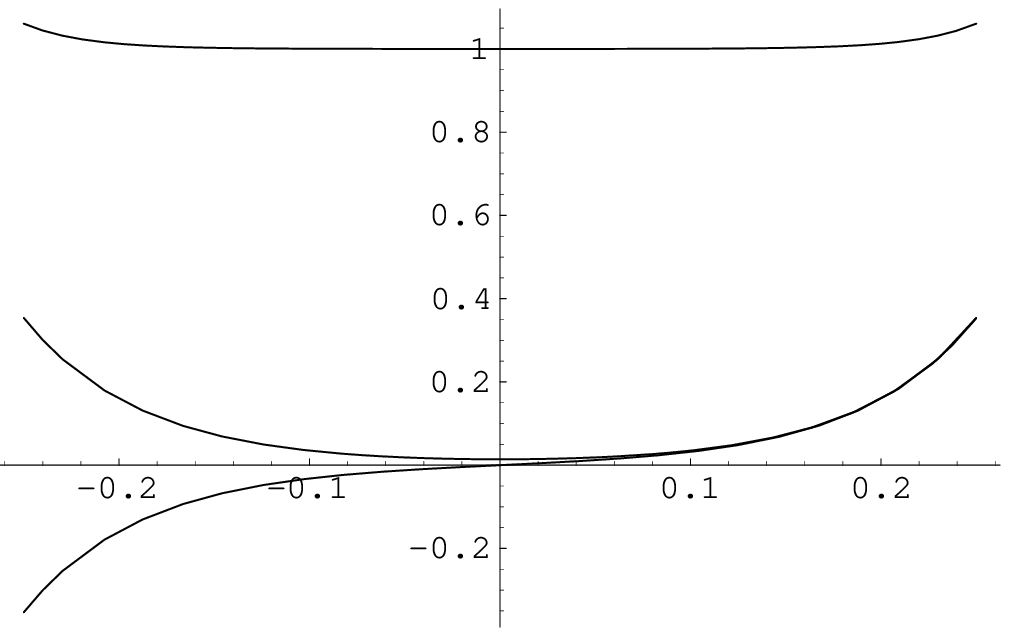}
\includegraphics{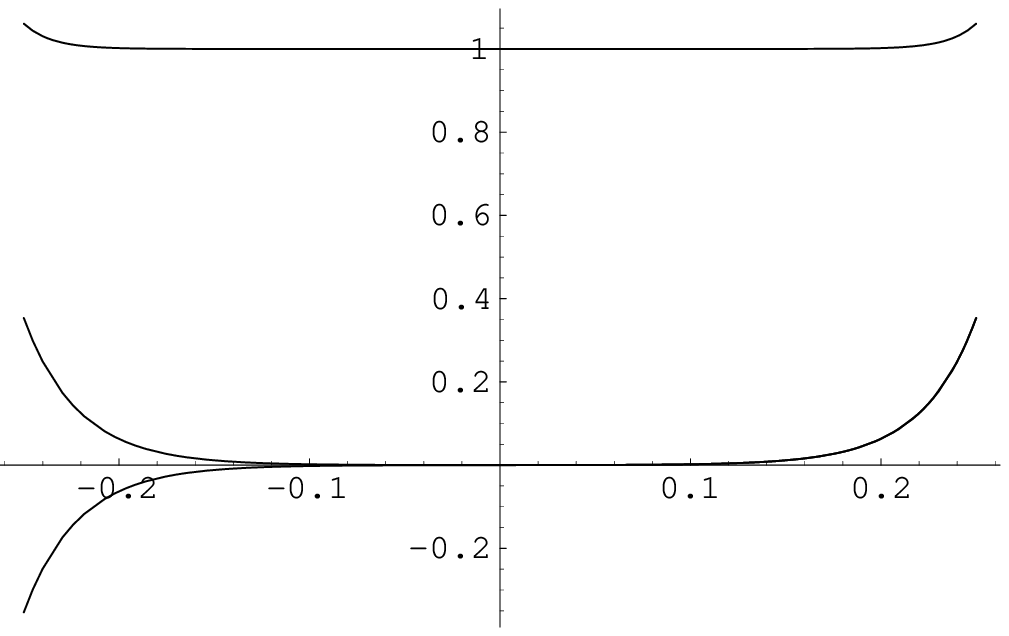}
\vskip2mm
\caption{Plots of $f_j(D(\k)z)$ for $z\in[-\quart,\quart]$ at $\k=1-10^{-4}$
(left) and $\k=1-10^{-8}$ (right), illustrating the approach to the point-like
limit $\k\to1$.}\label{fig:plfs}
\end{figure}

It is also interesting to inspect $\hat A_i(z)$ in the limit $\k\to1$ (or 
$\cD\to\infty$), to understand to which extent we retrieve the piecewize 
constant behavior of $\vec R(z;\vec x)$, on which the point-like limit is 
based. For this we plot $f_i(D(\k)z)$ in Fig.~\ref{fig:plfs}, which apart 
from fixed rotations and an overall factor $\cD$ would represent the 
constituent locations. Since $\cD(\k)/(\pi\rho^2)$ approaches $\sqrt{2}$ for 
$\k\to1$, it means we normalize the constituent locations with respect to 
height of the jumps in the Nahm data. At the impurities ($z=\pm1/4$) we 
therefore expect $f_i(D(\k)z)$ to go to a fixed value. The plotted cases,
$1-\k=10^{-4}$, $D(\k)=15.53$ (left) and $1-\k=10^{-8}$, $D(\k)=33.95$ 
(right), clearly demonstrate how in the bulk $f_3\to 1$ and $f_{1,2}\to0$, 
but that they differ at $z=\pm1/4$, to accommodate the discontinuities 
of the Nahm equation at the impurities. The cross-over from bulk behavior 
to the impurity values scales as $D^{-1}$, and is only absent for axially 
symmetric solutions.

\subsection{Extended structure}\label{sec:ext}
When $\vec R(z;\vec x)$ is not piecewize constant, it is not clear what plays
the role of the constituent locations. We therefore do expect some extended 
structure for $\k\neq1$. For this reason we now come back to analysing 
$\cV_m(\vec x)=(4\pi)^{-1}\Tr R_m^{-1}(z)$ in more detail. Quite remarkably, 
the discussion in Sect.~\ref{sec:cons} implies that the result for $\k=0$ 
can be obtained from that at $\k=1$. For this we may use the symmetry 
$c_2\leftrightarrow c_3$, which leaves $\cV_m(\vec x)$ invariant, apart from 
interchanging $x_2$ and $x_3$ ($\hat M$ is left unchanged when absorbing the 
rotation $\cR$ that interchanges $c_2$ and $c_3$ in $U$, see \refeq{defMhat}). 
With the definitions of $D=4\pi\cD$ and $\k$ in \refeq{Dkdef} one finds that 
this implies $\cD^2\to-\cD^2$ and $\k\to\k'$. Therefore,
\beq
\k=0:\quad\cV_m(\vec x)=\frac{\sqrt{2}\sqrt{|\vec x|^2-\cD^2+\sqrt{(|\vec 
x|^2-\cD^2)^2+4\cD^2x_2^2}}}{4\pi\sqrt{(|\vec x|^2-\cD^2)^2+4\cD^2x_2^2}}.
\label{eq:k0p}
\eeq
For this we use that the $\k=1$ result, $\cV_m(\vec x)=(4\pi|\vec x-\vec y|
)^{-1}+(4\pi|\vec x+\vec y|)^{-1}$, with $\vec y=\cD(0,0,1)$, can be rewritten 
in the form of \refeq{k0p} by changing the sign of $\cD^2$ and interchanging 
$x_2$ and $x_3$. In some sense we may say that the result for $\k=0$ describes 
point-like constituents that have moved into a complex direction\footnote{This 
was observed before in the monopole context~\cite{NahmM}. It is also 
worthwhile to point out that the axially symmetric monopole solutions 
discussed there can appear as such in the caloron context, as we read off 
for $\k=0$ from \refeq{bcI}. In this case we expect to find another class of 
axially symmetric caloron solutions, with $\Delta\vec a$ giving the symmetry
axes, but since we are more interested here in the case of well-separated 
monopole constituents, we did not analyse this in further detail.}. We note 
that \refeq{k0p} is singular on the ring $x_2=0$, $|\vec x|^2=\cD^2$, i.e. on 
the circle of radius $|\cD|$ in the $1,3$-plane. But there is more of a 
surprise: the $x_2$ derivative is discontinuous on the disk bounded by the 
ring (away from the disc the function is smooth and harmonic), $\cV_m(\vec x)
=(2\pi)^{-1}|\cD x_2|/(\cD^2-r^2)^{3/2}(1+\cO(x_2^2))$, with $r^2\equiv x_1^2
+x_3^2$. This implies indeed an extended structure, with singularities on the 
entire disk. Before showing how to deal with this singularity structure we 
consider general values of $\k$.

As we have seen in Sect.~\ref{sec:cons}, $\Tr R_m^{-1}(z)$ is conserved 
as a consequence of the Nahm equation. Using for $R_m^\pm(z)$ the Riccati 
equation, \refeq{Riccati}, the fact that $\Tr R_m^{-1}(z)=2\Tr(R_m^+(z)
+R_m^-(z))^{-1}$ is independent of $z$ imposes a severe constraint. 
We make use of this by expanding $R_m^\pm(z)$ as a Taylor series in $z$. 
The Taylor coefficients can be expressed in terms of the initial conditions 
$R_m^\pm(0)$, through the explicit solution of $\hat A_j(z)$, \refeq{fnahm} 
(we make use of the invariance under translations and rotations to put 
$\vec a=\vec 0$, $h=\ein_2$ and $\cR=\ein_3$, as in Sect.~\ref{sec:cons}).  
We thus obtain the Taylor series for $\Tr R_m^{-1}(z)$, of which all 
coefficients should vanish except for the 0th order. This gives a set 
of {\em algebraic} equations for the initial conditions, $R_m^\pm(0)$, 
encoded in $X_\mu$ and $\tilde X_\mu$ through
\beq
R_m(0)=\half(R_m^+(0)+R_m^-(0))\equiv\cD(X_0\ein_2+X_j\tau_j),\quad
\half(R_m^+(0)-R_m^-(0))\equiv\cD(\tilde X_0\ein_2+\tilde X_j\tau_j),
\eeq
Note that $\cD$ enters as an overall scale factor, cmp. \refeq{fnahm}, 
such that
\beq
\cV_m(\vec x)=\frac{1}{4\pi}\Tr R_m^{-1}(0)=\frac{1}{2\pi\cD}\tilde
\cV(\cD\vec x),\quad\tilde\cV=\half\Tr\left(\frac{1}{X_0+X_j\tau_j}\right)=
\frac{X_0}{X_0^2-X_j^2}.
\eeq
Dependence on the coordinates and on $\k$ is mostly left implicit. The system 
of equations is of course hugely overdetermined, and some amount of good 
fortune was required in that the first 11 orders in the Taylor expansion were 
sufficient to solve for $X_\mu$ and $\tilde X_\mu$. Quite remarkably these 
imply that $\tilde X_0=\tilde X_1=\tilde X_3=X_2=0$, which considerably 
simplifies the task of solving for the remaining 4 variables. We found that 
\beq
\delta\equiv X_0^2-X_1^2-X_3^2-\tilde X_2^2
\eeq
satisfies the cubic equation
\beqa
&&\hskip-9mm(2-\k^2)\k^4+4\k^2(x_1^2-x_3^2)-\k^4(3x_1^2-x_2^2-x_3^2)+
\left((2-\k^2)(3x_1^2-x_2^2+x_3^2)-4x_1^2\right)|\vec x|^2\nonumber\\ 
&&\hskip-5mm -|\vec x|^6-\left(\k^4+2\k^2(x_1^2-x_2^2-x_3^2)+4x_2^2+
|\vec x|^4\right)\delta+\left(\k^2-2+|\vec x|^2\right)\delta^2+\delta^3=0,
\eeqa
whereas $X_1/X_0$, $X_3/X_0$ and $\tilde X_2$ can be solved for in terms of 
$\delta$, 
\beqa
X_1/X_0&=&\frac{|\vec x|^4-\delta^2+2(\k'^2(\delta-2x_1^2)+x_1^2+x_2^2-x_3^2)+
         (2-\k^2)\k^2}{4x_1\k'\k^2}\nonumber\\
X_3/X_0&=&\frac{|\vec x|^4-\delta^2+2(\delta-2x_3^2-\k'^2(x_1^2-x_2^2-x_3^2))-
         (2-\k^2)\k^2}{4x_3\k^2}\nonumber\\
\tilde X_2&=&\frac{|\vec x|^4-\delta^2+2\k^2(x_1^2-x_2^2-x_3^2)+4x_2^2+
          \k^4}{4x_2\k'}.
\eeqa
Therefore, $\tilde\cV^2$ is a rational function of $\delta$, $\vec x$ 
and $k$,
\beq
\tilde\cV^2=\frac{1}{(\delta+\tilde X_2^2)(1-X_1^2X_0^{-2}-X_3^2X_0^{-2})},
\eeq
and the proper root of the cubic equation for $\delta$ to use is fixed by 
$\tilde\cV\to1/|\vec x|$. Indeed, the asymptotic expansion for $\delta$,
\beq
\delta=|\vec x|^2\left(1+\frac{2x_2^2 +\k^2(x_1^2-x_2^2-x_3^2)}{|\vec x|^4}
+\ldots\right),
\eeq
reproduces the multipole expansion of $\tilde\cV$. This was verified to the 21 
orders given in \refeq{m20}. On general grounds it can be argued, since 
$\delta$ satisfies a cubic equation, that $\tilde\cV^2$ has to satisfy a 
cubic equation as well. Its coefficients (polynomials in $\k$ and $\vec x$) 
are somewhat lengthy, and therefore not reproduced here, in part because
we will shortly present an exact integral equation for $\cV_m(\vec x)$, valid
for any $\k$. The exact results for $\k=0$ and $\k=1$ is most easily recovered
from the cubic equation for $\tilde \cV^2$, but the integral representation
will be valid for these two cases as well.

Like for $\k=0$, we find that $\cV_m(\vec x)$ is harmonic everywhere except 
on a disk, now bounded by an ellipse with major and minor axes $\cD$, resp. 
$\k'\cD$. On this disk the function vanishes, satisfying in the direction 
perpendicular to the disk the expansion $\cV_m(\vec x)=(2\pi)^{-1}\k'|\cD 
x_2| /(\cD^2\k'^2-x_1^2-\k'^2x_3^2)^{3/2}(1+\cO(x_2^2))$ (cmp. the 
discussion above, for $\k=0$). By introducing the 
``polar" coordinates $(x_1,x_3)=(\k'r\cos\vphi,r\sin\vphi)$, in terms of which 
the ellipse at the boundary of the disk is characterized by $r=\cD$, this can 
be written as $\cV_m(\vec x)=(2\pi\k')^{-1}|\cD x_2|/(\cD^2-r^2)^{3/2}(1+
\cO(x_2^2))$.  Care is required to deal correctly with the behavior at the 
edge of the disk when computing the Laplacian. We find
\beq
-\partial_i^2\cV_m(\vec x)=-\delta(x_2)\frac{\cD}{\pi r\k'}\frac{\partial}{
\partial r}\frac{\theta(\cD-r)}{\sqrt{\cD^2-r^2}},\label{eq:chdist}
\eeq
with $\theta(\cD-r)$ the step function, giving the following integral 
representation 
\beq
\cV_m(\vec x)=\frac{1}{2\pi|\vec x|}+\frac{\cD}{4\pi^2}\int_0^{2\pi}
\!\!\!\!\!d\vphi\int_0^\cD\!\!\!\frac{dr}{\sqrt{\cD^2-r^2}}\partial_r
\frac{1}{\sqrt{(x_1-\k'r\cos\vphi)^2+(x_3-r\sin\vphi)^2+x_2^2}}.\label{eq:res}
\eeq
Note that $\k'$, appearing in the denominator of \refeq{chdist}, cancels due to
the change of variables to ``polar" coordinates. By numerical evaluation, we 
checked this formula against the exact results. It gives us confidence that 
we interpreted the singularity structure correctly. Taking an arbitrary test 
function $f(\vec x)$ we find
\beq
\cN(f)\equiv-\int f(\vec x)\partial_i^2\cV_m(\vec x)d^3x=2f(\vec 0)+
\frac{\cD}{\pi}\int_0^{2\pi}d\vphi\int_0^\cD dr~\frac{\partial_r f
(\k'r\cos\vphi,0,r\sin\vphi)}{\sqrt{\cD^2-r^2}}.\label{eq:distest}
\eeq
The integral over $r$ is well defined for any $\k'$ and can be used to check 
the correct normalization for the integrated zero-mode density, $\cN(1)=2$. 
\refeq{distest} is also particularly convenient for studying the limit 
$\k'\to0$. Using the fact that $\cN(f)$  is even in $\k'$, we may write 
$\cN(f)=\cN_0(f)+\k'^2 \cN_2(f)+\cO(\k'^4)$, with in particular
\beq
\cN_0(f)=2f(\vec 0)+\frac{\cD}{\pi}\int_{-\cD}^\cD dy\int_{_{-\sqrt{\cD^2-y^2}}
}^{^{\sqrt{\cD^2-y^2}}}dx~\frac{y\partial_yf(0,0,y)}{(x^2+y^2)
\sqrt{\cD^2-x^2-y^2}},
\eeq
reintroducing cartesian coordinates 
$x=r\cos\vphi$ and $y=r\sin\vphi$. The integral over $x$ is easily 
performed and we find for $\cN_0(f)$
\beq
\cN_0(f)=2f(\vec 0)+\int_{-\cD}^\cD dy~{\rm sign}(y)\partial_yf(0,0,y)=
f(0,0,\cD)+f(0,0,-\cD),
\eeq
whereas $\cN_2(f)$ gets contributions from $f$ on the line between $(0,0,\cD)$ 
and $(0,0,-\cD)$. Therefore, we conclude that in the limit $\k \to1$ two 
point-like constituents are found, and that this limit is approached in a 
smooth way (despite the behavior observed in Fig.~\ref{fig:plfs}). 
\begin{figure}[htb]
\vskip5.1cm
\includegraphics{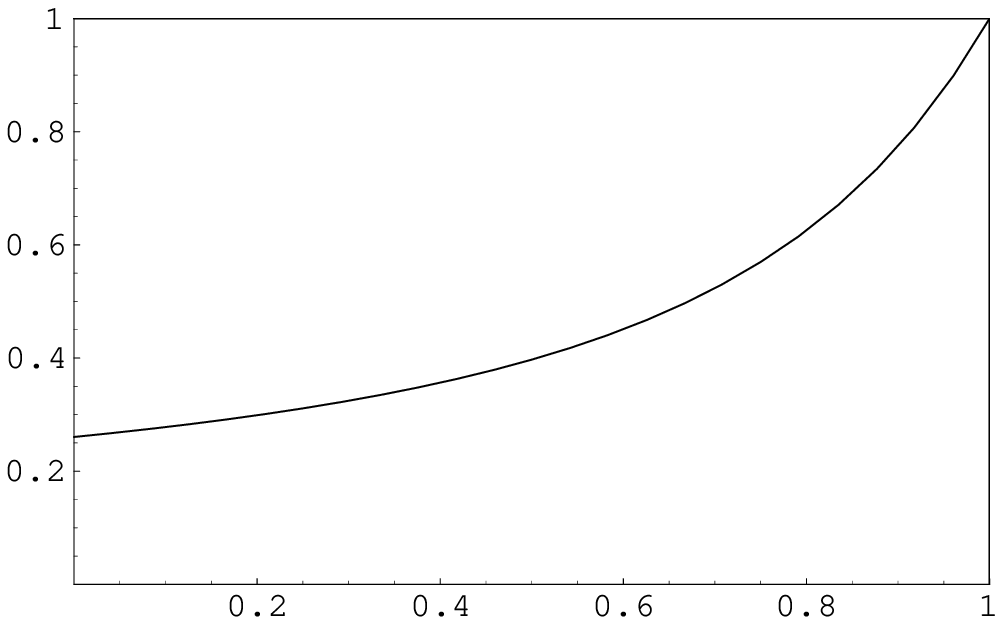}
\caption{$\cN(f)$, \refeq{distest}, as a function of $\k^2$ for $f(\vec x)=
\exp[-10(x_1^2+x_2^2+(x_3-1)^2)]$ and $\cD=1$.}\label{fig:testf}
\end{figure}
As an illustration we plot in Fig.~\ref{fig:testf} $\cN(f)$ as a function of 
$\k^2$ for a Gaussian centered at one of the would-be constituent locations, 
where it takes the value 1. We see that indeed $\cN(f)$ reaches 1 (linear 
in $\k'^2$) for $\k\to1$. We also recall that \refeq{kpasym} implies the 
point-like limit is reached exponentially in the constituent monopole 
separation ($2\cD$).

For any $\k\neq1$ the core has an extended structure in the high temperature 
limit. One might expect it to be extended along a line, since $\hat A_j(z)$ 
depends on a single parameter. With $\vec R(z;\vec x)$ not constant, the 
Riccati equation apparently ``smears out" the core, but suprisingly only to a
disk. Note that inside this disk the singular zero-mode density is negative. 
To guarantee the proper normalization, this is compensated by the singular 
behavior at the ellipse which forms the edge of the disk. It should be noted, 
however, that the singularity structure obtained in the high temperature limit 
will only tell us to which region the core will be restricted. We will need 
to resolve the non-abelian field components inside the core to understand how 
this limiting behavior comes about. 

\section{Summary and Discussion}\label{sec:disc}

We studied the zero-modes of higher charge calorons, in particular in the 
far field limit. In this limit one considers the regions outside the cores 
of the constituents monopoles, where the gauge field is abelian. The
constituent monopoles are most prominent when they all have a non-zero mass,
implying a non-trivial value of the holonomy. Since the holonomy can be seen 
as a background Polyakov loop, these calorons are therefore more relevant 
for the confined phase, where the average Polyakov loop is non-trivial. 
Lattice evidence for the relevance of these configurations is steadily 
increasing~\cite{Ilg2,Gattp}, and much analytic understanding of the charge 1 
calorons has been gained~\cite{KvB}. These reveal $n$ constituents for 
$SU(n)$, which when well separated become static BPS monopoles. To localize 
these monopoles it is useful to consider the high temperature limit, for 
which the masses go to infinity and the non-abelian cores shrink to zero 
size. We note that the high temperature limit is equivalent to the limit  
of infinite Higgs vacuum expectation value (recall that $A_0$ plays the role 
of the Higgs field in the adjoint representation). Since also chiral fermions 
in a caloron background generically have a mass, likewise going to infinity 
in the high temperature limit, one finds these zero-modes to be localized 
entirely to the non-abelian cores. Indeed for charge 1, the zero-mode density 
becomes a delta function at the location of one of the constituent monopoles. 
Which one of them, is uniquely determined by the holonomy, and the phase up 
to which the zero-mode is chosen to be periodic in the imaginary time 
direction. We generalize away from anti-periodic boundary conditions since 
this freedom in choosing the phase allows us to control on which constituent 
the zero-mode localizes. It thus allows us to probe the underlying gauge 
field configuration, see Fig.~\ref{fig:zmcycle}, as has been convincingly 
demonstrated in Monte Carlo studies as well~\cite{Gattp}.

For higher charge $SU(n)$ calorons, $k>1$, there will be $kn$ constituent
monopoles, or more precisely $k$ constituent monopoles of given type. Within 
each type the magnetic charge associated to the embedding in $U^{n-1}(1)$, 
and the mass is fixed. In a previous paper~\cite{BrvB} we constructed 
axially symmetric solutions, which shared many of the properties of the 
charge 1 solutions. In particular the high temperature limit gave 
point-like constituents and we have verified here that the zero-mode 
density indeed localizes to these constituents in the expected way, see 
Fig.~\ref{fig:zmlocal}. This, however, is not expected to hold for higher 
charge calorons in general. The reason is that one constructs these solutions 
with the help of the Nahm equation, described by a dual $U(k)$ gauge field 
$\hat A_j(z)$ on the circle, with singularities at the eigenvalues $\mu_m$ 
of the logarithm of the holonomy. In case of the axially symmetric solutions 
of Ref.~\cite{BrvB} this Nahm gauge field is actually piecewize constant 
(apart from some trivial $U(k)$ gauge rotation), and its value on the interval 
$z\in[\mu_m,\mu_{m+1}]$ directly determines the locations of the type $m$ 
monopole constituents. However, in general the Nahm gauge field will not be 
piecewize constant. In that case it remained to be seen if the constituent 
locations are smeared out over all values of $\hat A(z)$, or worse.

Indeed in this paper we have found for charge 2 calorons in $SU(2)$ that in 
general the location of the constituents is smeared out over a disk. To be
more precise, we have shown that the zero-mode density vanishes everywhere,
except on a disk that is bounded by an ellipse. We wish to stress that
this result is found by solving the Nahm equation on an interval, without
imposing the boundary conditions. In other words, a disk is described by 
the parameters of the solution for each interval, that is for each monopole 
type, namely the center of mass, the orientation, a shape and a scale 
parameter (eight in total). These should also be the building blocks for 
higher gauge groups $SU(n)$ since their dual descriptions differ only in 
the number of intervals. Moreover, only the boundary conditions distinguish 
between calorons and monopole solutions, and we believe our result is of 
value in the context of the latter as well.

In particular we want to draw attention to the fact that (the trace of) the 
chiral fermion zero-mode density in the high temperature limit is the Laplacian
of a function $-\cV_m(\vec x)$ that is determined in terms of the Nahm gauge 
field and turns out to be a constant of the motion with respect to the Nahm 
equation. We have verified this property by computing the multipole expansion 
of $\cV_m(\vec x)$ to a high order, but we expect this can be proven from the 
integrability of the Nahm equations, an interesting problem to be pursued in 
the future. The fact that $\cV_m(\vec x)$ is conserved is a powerful result 
indeed, since it allowed us to calculate this function exactly, on which we 
base the findings mentioned above. 

This makes us conjecture that the cores of the constituent monopoles are in 
general extended, collapsing to the disk in the high temperature limit. In 
itself this is not surprising, since one knows from the study of monopoles, 
when two are closer together than the size of their core, they show an extended 
structure, for charge 2 indeed in the shape of a doughnut~\cite{MonRing}. It 
should also be noted that we found, when approaching the disk from above with 
a test function smaller than the ring, the resulting zero-mode density can be 
negative. There are two reasons why we should not be too worried about this. 
If the core would collapse to points, the proper normalization of the 
zero-modes requires the contribution from the singularity to be quantized, 
giving a delta function of unit strength. When the core is extended, we have 
no such constraint. In addition, our equations are derived by ignoring the 
exponentially small terms that occur in the core. Therefore our result is 
only valid outside the core, where we indeed find the zero-mode density to 
vanish. This means the cores have to lie within the disk, whereas the only 
thing we can say about their contributions is that they have to integrate to 
2 (since $-\partial_i^2\cV_m(\vec x)$ adds the two zero-mode densities). 
Although this may seem to resemble a singularity like the Dirac string, it 
is more subtle than that, because only like-charge monopoles are involved 
here. Nevertheless, it does show that the non-abelian fields inside the 
core have an intricate behavior. 

It will therefore be interesting in the future to try and get access to the 
gauge field and to see if the field strength shows further localization within 
the disks we have found on the basis of the zero-modes. Another useful object 
is the Polyakov loop because it traces the constituent monopoles~\cite{MTAP}
and is able to reveal extended structures in the context of Abelian 
projection~\cite{AbPr,Ford}. We believe to have gained sufficient information 
to get access to the solutions at finite temperature, with a fully resolved 
core to answer these questions, if necessary by numerical means.

However, from the physical point of view, the most important result of this
paper is our demonstration that for well separated constituents, when the 
scale parameter $D$ is large, the shape of the caloron zero-mode density leads 
to point-like constituents, i.e. the shape parameter $\k$ approaches 1. It 
does so exponentially in $D$ and we have shown no structure is left on the 
disk bounded by the ellipse, collapsing to a line in this limit. Any trace 
left over on this line is proportional to $1-\k^2$, and therefore vanishes 
exponentially in $D$. This comes about through the boundary conditions 
$\hat A(z)$ has to satisfy at the ``impurities" $\mu_m$, relating the size 
and shape parameters, $D$ and $\k$. We leave it to a future publication to 
more fully describe the moduli space of solutions, solving for these 
constraints. Also here some remarkable simplifications seem to occur, 
related to the integrability of the Nahm equations. 

The results of this paper therefore provide one further step in establishing
that non-abelian gauge field configurations can be described on a large 
distance scale in terms of abelian monopoles. Of course it is only a small
step, because we use exact self-dual solutions to establish these results.
This has been in part because we discovered earlier~\cite{BrvB}, somewhat to 
our surprise, that it is far from trivial to write down superpositions of 
these monopole fields without having visible Dirac strings all over the 
place, that would carry too much energy for comfort. In part this is due to 
the crudeness involved in the superposition, instead requiring a fine-tuning 
of the non-abelian tails with the exponential components in the abelian gauge 
field (to properly absorb the return flux). It has been this problem to deal 
with Dirac strings that for so long has hampered attempts to describe an 
interacting theory of magnetic monopoles~\cite{Zwan}. 

In the light of this it would of course be welcome if more lattice studies
are performed to get a handle on the dynamics of these constituent monopoles. 
It should be pointed out that instantons larger than $\beta$ will no longer 
reveal themselves as lumps of size $\rho$. Rather there is a transition region 
beyond which $\rho$ should be interpreted to set the inter-constituent 
distance (typically of order $\pi\rho^2/\beta$), whereas the size of the lumps 
is in this region set by the mass of the constituent monopoles. This may lead 
to a natural infrared cutoff in the size distribution of instantons, provided 
by the temperature (rather than the spatial volume). Because of this it may be 
worthwhile to reinvestigate the issue of the instanton size distribution, also 
with the criticism presented in Ref.~\cite{Horv} in mind.

\section*{Appendix A}

We will derive in this appendix the zero-mode limit. Our starting point is the 
explicit expression for the Green's function, \refeq{fdf}. It is useful to 
note that this Green's function satisfies $f_x(z,z')=f_x^\dagger(z',z)$. To 
show that \refeq{fdf} indeed respects this relation, one can use 
(see \refeq{Wdef})
\beqa
&&\pmatrix{0&-\ein_k\cr\ein_k&0\cr}W^\dagger(z,z')\pmatrix{0&\ein_k\cr-\ein_k
&0\cr}=W^{-1}(z,z'),\label{eq:Wdagger}\\&&\pmatrix{0&-\ein_k\cr\ein_k&0\cr}
(\ein_{2k}-\cF_{z_0}^\dagger)^{-1}\pmatrix{0&\ein_k\cr-\ein_k&0\cr}=(
\ein_{2k}-\cF_{z_0}^{-1})^{-1}=\ein_{2k}-(\ein_{2k}-\cF_{z_0})^{-1}.\nonumber
\eeqa
We will take $z_0=\mu_m+0$, which leads to the following decomposition of 
$\cF_{\mu_m}$ in terms of contributions coming from the impurities ($T_m$) 
and from the propagation between the impurities ($H_m$),
\beqa
&&\cF_{\mu_m}=T_mH_{m-1}\cdots T_2H_1T_1\hat g^\dagger(1)H_n T_nH_{n-1}
\cdots T_{m+1}H_m,\nonumber\\&& T_m=\exp\pmatrix{0&0\cr 2\pi S_m&0\cr},\quad 
H_m=\hat W_m(\mu_{m+1})F_m(\mu_{m+1})\hat W_m^{-1}(\mu_m).\label{eq:THs}
\eeqa
Note that by definition $F_m(\mu_m)=\ein_k$. It is convenient to absorb 
the algebraic contributions coming from $\hat W_m$ in the ``impurity" 
contributions $T_m$.
\beqa
\Theta_m&=&\pmatrix{\theta^m_{++}&\theta^m_{+-}\cr\theta^m_{-+}&\theta^m_{--}
\cr}\equiv\hat W_m^{-1}(\mu_m)T_m\hat W_{m-1}(\mu_m)\label{eq:Theta}\\
&=&\half R_m^{-1}(\mu_m)\pmatrix{
R_m^-(\mu_m)+R_{m-1}^+(\mu_m)+S_m&R_m^-(\mu_m)-R_{m-1}^-(\mu_m)+S_m\cr
R_m^+(\mu_m)-R_{m-1}^+(\mu_m)-S_m&R_m^+(\mu_m)+R_{m-1}^-(\mu_m)-S_m\cr}.
\nonumber
\eeqa
The following identities are noteworthy
\beq
\theta^m_{+\pm}+\theta_{-\pm}^m=\ein_k,\quad\theta^m_{\pm+}-\theta_{\pm-}^m=
\pm R_m^{-1}(\mu_m)R_{m-1}(\mu_m),\quad 2R_m(\mu_m)\theta^m_{++}=\Sigma_m,
\label{eq:thetas}
\eeq
as well as the fact that $\Theta^{-1}_m=\hat W_{m-1}^{-1}(\mu_m)T_m^{-1}
\hat W_m(\mu_m)$ can be computed explicitly
\beq
\Theta^{-1}_m=\pmatrix{\theta_m^{++}&\theta_m^{+-}\cr\theta_m^{-+}&
\theta_m^{--}\cr}=R_{m-1}^{-1}(\mu_m)R_m(\mu_m)\pmatrix{\theta^m_{--}&
-\theta^m_{+-}\cr-\theta^m_{-+}&\theta^m_{++}\cr}.\label{eq:Thetainv}
\eeq
Finally, introducing 
\beq
\hat\cF_{\mu_m}=W^{-1}_m(\mu_m)\cF_{\mu_m}W_m(\mu_m)=\Theta_mF_{m-1}
\cdots\Theta_2F_1\hat g^\dagger(1)\Theta_nF_{n-1}\cdots\Theta_{m+1}F_m,
\label{eq:cFhat}
\eeq
we find for $\mu_m\leq z'\leq z\leq\mu_{m+1}$ (cmp. \refeq{fdf}) 
\beq
f_x(z,z')=-4\pi^2\pmatrix{\ein_k\cr0\cr}^tW_m(z)(\ein_{2k}-\hat
\cF_{\mu_m})^{-1}W_m^{-1}(z')\pmatrix{0\cr\ein_k\cr}.\label{eq:farf}
\eeq
It will be useful to write $\hat\cF_{\mu_m}=F_m^{-1}\Theta_{m+1}^{-1}LK
\Theta_{m+1}F_m$, with $L\equiv\Theta_{m+1}F_m\Theta_m$, because 
$K\equiv F_{m-1}\Theta_{m-1}\cdots\Theta_1\hat g^\dagger(1)F_n\Theta_n
\cdots\Theta_{m+2}F_{m+1}$ contains the exponential factors in terms of 
which we can take the zero-mode limit. 

With $(\ein_{2k}-\hat\cF_{\mu_m})^{-1}=F_m^{-1}\Theta_{m+1}^{-1}(1-LK)^{-1}
\Theta_{m+1}F_m$, we reduce the problem to approximating $(\ein_{2k}-LK)^{-1}$.
For this it is convenient to write $LK\equiv\hat L\hat K+\tilde L\tilde K$, 
with
\beq
\hat K\equiv\pmatrix{K_{++}&K_{+-}\cr 0&\ein_k\cr},\quad
\tilde K\equiv\pmatrix{0&0\cr K_{-+}&K_{--}\cr},\quad
\hat L\equiv\pmatrix{L_{++}&0\cr L_{-+}&0\cr},\quad
\tilde L\equiv\pmatrix{0&L_{+-}\cr0&L_{--}\cr}.
\eeq
after which we find
\beq
(\ein_{2k}-LK)^{-1}=\hat K^{-1}\left(\hat K^{-1}-\hat L
-\tilde L\tilde K\hat K^{-1}\right)^{-1}.
\eeq
As we will show next, the advantage of all this is that terms containing
$K_{\pm\pm}$ are of the form $K_{++}^{-1}$, $K_{++}^{-1}K_{+-}$, $K_{-+}
K_{++}^{-1}$ or $(K^{--})^{-1}\equiv K_{--}-K_{-+}K_{++}^{-1}K_{+-}$ and
that these are all exponentially decaying. For the first three this is 
easily seen using that $K$ is of the form $F\Theta F\Theta\cdots F\Theta F$, 
whereas for the last term we recall a well-known formula for the inverse of 
a $2\times 2$ matrix with as entries ($k\times k$) matrices
\beq
K^{-1}=\pmatrix{K_{++}&K_{+-}\cr K_{-+}&K_{--}\cr}^{-1}=
\pmatrix{(K_{++}-K_{+-}K^{-1}_{--}K_{-+})^{-1}&
         (K_{-+}-K_{--}K^{-1}_{+-}K_{++})^{-1}\cr
         (K_{+-}-K_{++}K^{-1}_{-+}K_{--})^{-1}&
         (K_{--}-K_{-+}K^{-1}_{++}K_{+-})^{-1}}.\label{eq:Kinv}
\eeq
From this we find that $(K^{--})^{-1}=(K^{-1})_{--}$ (hence the upper 
indices). With $K^{-1}$ having the form $F^{-1}\Theta^{-1}F^{-1}\Theta^{-1}
\cdots F^{-1}\Theta^{-1} F^{-1}$, which interchanges the role of $f^+$ and 
$f^-$, we conclude that $K^{--}$ behaves as $K_{++}$ and that therefore 
$(K^{--})^{-1}$ is exponentially decaying as well. Using 
\beq
\hat K^{-1}=\pmatrix{K^{-1}_{++}&-K^{-1}_{++}K_{+-}\cr 0&\ein_k\cr},\quad
\tilde K\hat K^{-1}=\pmatrix{0&0\cr K_{-+}K^{-1}_{++}&(K^{--})^{-1}\cr}
\eeq
and neglecting the exponentially decaying terms, we find the simple result
\beq
(\ein_{2k}-LK)^{-1}\to\pmatrix{0&0\cr0&\ein_k}\pmatrix{-L_{++}&0\cr
-L_{-+}&\ein_k}^{-1}=\pmatrix{0&0\cr -L_{-+}L_{++}^{-1}&\ein_k},\label{eq:ffl}
\eeq
or including subleading terms
\beq
(\ein_{2k}-LK)^{-1}=\pmatrix{\cO(K_{++}^{-1})&\cO(K_{++}^{-1})\cr -L_{-+}
L_{++}^{-1}+\cO(X)&\ein_k+\cO(X)},\label{eq:subffl}
\eeq
where $\cO(X)\equiv\cO(K_{++}^{-1}K_{+-})+\cO(K_{-+}K_{++}^{-1})+
\cO((K^{--})^{-1})$.

Using \refeq{ffl} and $\Theta_{m+1}F_m=L\Theta_m^{-1}$ we find
\beq
f^{\zm}_x(z,z')=\pi\pmatrix{\ein_k\cr\ein_k\cr}^t F_m(z)F_m^{-1}
\Theta_{m+1}^{-1}\pmatrix{0&0\cr -L_{-+}L_{++}^{-1}&\ein_k\cr}L\Theta_m^{-1}
F_m(z')\pmatrix{-\ein_k\cr\hphantom{-}\ein_k\cr}R^{-1}_m(z'),
\eeq
This can be simplified further using $L^{--}\equiv(L^{-1})_{--}=(L_{--}-
L_{-+}L^{-1}_{++}L_{+-})^{-1}$ (cmp. \refeq{Kinv}), such that 
\beq
f^{\zm}_x(z,z')=\pi\pmatrix{f_m^+(z)\cr f_m^-(z)\cr}^tF_m^{-1}
\Theta_{m+1}^{-1}\pmatrix{0&0\cr0&(L^{--})^{-1}\cr}\Theta_m^{-1}
\pmatrix{-f_m^+(z')^{-1}\cr\hphantom{-}f_m^-(z')^{-1}\cr}R^{-1}_m(z').
\eeq
With Eqs.~(\ref{eq:thetas},\ref{eq:Thetainv}), noting that $Z_m^-=-
\theta_m^{+-}(\theta_m^{--})^{-1}$ and $Z_m^+=(\theta_m^{--})^{-1}
\theta_m^{-+}$ (see \refeq{Zedef}), we find after some algebra the relatively 
simple result given in \refeq{zmlim}. We note that it is not directly obvious 
that $f^\zm_x(z,z')=f_x^\zm(z',z)^\dagger$. Nevertheless, this is guaranteed 
to be true from the fact that the exact Green's function respects this 
property. All we wish to mention here, is that \refeq{Wdagger} implies rather 
non-trivial relations involving $f_m^\pm(z)^\dagger$ and $R_m^\pm(z)^\dagger$, 
which could be used to explicitly verify that $f^\zm_x(z,z')=
f_x^\zm(z',z)^\dagger$. 

\section*{Appendix B}

Using the invariance under a one-parameter set of rotations around $\hat x$ 
for $M$, \refeq{defM}, or equivalently around $(1,0,0)$ for $\hat M(\hat x)$, 
\refeq{defMhat}, we can for charge 2 express the multipole expansion of 
$\cV_m(\vec x)$ in the following 4 independent parameters,
\beqa
&&\!\!\!p\equiv\frac{3}{2}\hat M_{11}(\hat x),\quad w^2\equiv\hat M_{12}^2(\hat
x)+\hat M_{13}^2(\hat x),\quad q^2\equiv\frac{1}{8}\left(\hat M_{11}(\hat x)+
2\hat M_{22}(\hat x)\right)^2\!\!+\frac{1}{2}\hat M_{23}^2(\hat x),\nonumber\\
&&\!\!\!\!s^3\equiv4\hat M_{12}(\hat x)\hat M_{13}(\hat x)\hat M_{23}(\hat x)+
\left(\hat M_{12}^2(\hat x)-\hat M_{13}^2(\hat x)\right)\left(\hat M_{11}
(\hat x)+2\hat M_{22}(\hat x)\right),
\eeqa
where 
$|\vec x|^2p$, $|\vec x|^4w^2$, $|\vec x|^4q^2$ and $|\vec x|^6s^3$ can be
written as monomials in $\vec x$. This choice has the particular advantage 
that for charge 2 the following remarkably simple form can be used 
\beq
\cV_m(\vec x)=\frac{1}{2\pi}\sum_{n=0}\frac{a_n}{|\vec x|^{2n+1}},\quad
\partial_pa_n=na_{n-1},\label{eq:mpol}
\eeq
checked to order $|\vec x|^{-21}$, but likely to be true to all orders.
Therefore, the result to this order can be read off from the $l=2n=20$ 
multipole coefficient 
\beqa
a_{10}&=&\!\!p^{10}\!+45p^8(q^2\!-2w^2)+180p^7s^3\!+315p^6(q^4\!-8q^2w^2\!+
  4w^4)+630p^5s^3(3q^2\!-4w^2)\nonumber\\ 
&+&\!\!\frac{525}{2}p^4(2q^6-36q^4w^2+60q^2w^4-16w^6+3s^6)+
  3150p^3s^3(q^4-4q^2w^2+2w^4)\nonumber\\ 
&+&\!\!\frac{315}{8}p^2(5q^8-160q^6w^2+560q^4w^4-448q^2w^6+40q^2s^6+
  80w^8-48w^2s^6)\nonumber\\
&+&\!\!\frac{105}{2}ps^3(15q^6-120q^4w^2+168q^2w^4-48w^6+2s^6)+ 
  \frac{63}{8}\Bigl(q^{10}-50q^8w^2\nonumber\\
&&\quad+300q^6w^4+5q^4(5s^6-96w^6)+80q^2(3w^8-w^2s^6)+40w^4s^6-32w^{10}\Bigr).
\label{eq:m20}
\eeqa

\section*{Acknowledgements}

We thank Conor Houghton for sharing his insights and unpublished notes, 
as well as Andreas Wipf, David Adams and in particular Chris Ford for 
discussions. We also thank Michael M\"uller-Preussker, Michael Ilgenfritz, 
Boris Martemyanov, Stanislav Shcheredin and Christof Gattringer for 
stimulating discussions concerning calorons with non-trivial holonomy 
on the lattice. The research of FB is supported by FOM.

\end{document}